\PassOptionsToPackage{dvipsnames}{xcolor}
\documentclass[onecolumn,floatfix]{aastex631}
\usepackage{amsmath}
\usepackage{amsfonts}
\usepackage{graphicx}
\usepackage{xcolor}
\usepackage{parskip}
\hypersetup{hypertexnames=false}

\makeatletter
\long\def\@makefntext#1{%
  \sbox\z@{\mbox{}\hspace*{3mm}\@makefnmark~}%
  \hsize=\dimexpr\columnwidth-\wd\z@\relax
  \mbox{}\hspace*{3mm}\@makefnmark~#1%
  \hsize=\columnwidth}
\makeatother

\makeatletter
\long\def\frontmatter@abstractheading{%
 \begingroup
  \centering
\ifmodern\else\hskip-0.05\textwidth \fi \abstractname
  \vskip 1mm
  \par
 \endgroup
}%
\makeatother

\DeclareMathOperator{\solve}{solve}
\DeclareMathOperator{\svd}{svd}
\DeclareMathOperator{\eig}{eig}
\DeclareMathOperator*{\argmin}{argmin}

\DeclareMathOperator*{\score}{score}

\begin{document}

\title{Robust Heteroskedastic Matrix Factorization: \\
  A Generalization of PCA that Flags Outliers and Handles Missing Data}
\author[0000-0001-7641-5235]{Thomas Hilder}
\affiliation{School of Physics and Astronomy, Monash University VIC 3800, Australia}

\author[0000-0003-2866-9403]{David W. Hogg}
\affiliation{Center for Cosmology and Particle Physics, Department of Physics, New York University, 726~Broadway, New~York, NY~10012,~USA}
\affiliation{Max-Planck-Institut f\"ur Astronomie, K\"onigstuhl 17, D-69117 Heidelberg, Germany}
\affiliation{Center for Computational Astrophysics, Flatiron Institute, 162~Fifth~Ave., New~York, NY~10010,~USA}

\author[0000-0003-0174-0564]{Andrew R.\ Casey}
\affiliation{School of Physics and Astronomy, Monash University VIC 3800, Australia}
\affiliation{Center for Computational Astrophysics, Flatiron Institute, 162~Fifth~Ave., New~York, NY~10010,~USA}

\author[0000-0003-4996-9069]{Hans-Walter Rix}
\affiliation{Max-Planck-Institut f\"ur Astronomie, K\"onigstuhl 17, D-69117 Heidelberg, Germany}

\begin{abstract}\noindent\hsize=\textwidth\leftskip=0.05\textwidth\rightskip=\leftskip 
  We present Robust Heteroskedastic Matrix Factorization (RHMF), a generalization of Principal Component Analysis (PCA) that is robust to outliers, handles per-feature uncertainties and missing data, and automatically flags per-feature and per-object anomalies.
  RHMF is useful both in recovering a low-dimensional embedding unspoiled by bad data or anomalies, and in identifying those anomalies.
  It utilises an iterative reweighting algorithm that implicitly maximizes a Student-t likelihood.
  This admits an equivalent probabilistic interpretation as fitting a hierarchical model with per-data-point latent variances.
  We deliver a fast \texttt{JAX} implementation, \texttt{Robusta-HMF}, and practical guidance for users. We demonstrate the ability of the model to identify and mitigate outliers of different classes.
  Identification accuracy is contingent on the choice of hyperparameters, but we show that these can be set reliably by cross-validation.
  We also apply RHMF to RVS spectra from Gaia DR3 to find  stars that are strange relative to their neighbors in color-magnitude space, or strange relative to all other stars.
  We highlight specific examples, including a known binary hosting a Be star, blue supergiants, Mira variables, stars with anomalous abundances, and M-dwarfs with subtle emission in the Ca II triplet lines, indicative of accretion or magnetic activity, which would not be obvious to identify by eye.
  \end{abstract}

\section{Introduction}

Many high-dimensional datasets in science and engineering naturally lie on low-dimensional manifolds. For example, most stellar spectra are well described by just a handful of physical parameters---effective temperature, surface gravity, metallicity, $\alpha$-element enhancement, and perhaps projected rotation and a continuum shape---that account for most of the variation across thousands of pixels of a given spectrum, for most stars of most types. This is both an expectation of stellar atmosphere theory \citep{hubeny2014} and an empirical finding from data-driven analyses of large spectroscopic surveys \citep[e.g.][]{ness2015,casey2016,price-jones2018}. A central task of modern stellar spectroscopy, and more broadly any survey dealing with high-dimensional data, is to identify this manifold from the data, represent it compactly, and place every observed  object on it. In doing so, we can improve any subsequent label estimation, de-noise  observations, and find astrophysical outliers: often the most scientifically interesting objects.

This work tackles two practical problems that arise in pursuit of this goal. The first is manifold recovery and projection: How can we learn a low-dimensional basis from a collection of spectra, and for each spectrum, find the coefficients that reconstruct it from that basis? A good embedding of this type is useful in equal measure for label estimation, denoising, and compression. The hard part is not the linear algebra, but rather the data. Real surveys are heteroskedastic, with missing segments, corrupted pixels, and innumerable other peculiarities. Under these conditions, standard methods to find a low-dimensional embedding will often fail or perform poorly.

The second problem is identifying ``outliers'' and handling them appropriately. With a good low-dimensional manifold, objects (stars) and features (pixels) that are \emph{not} represented on the manifold are of particular interest. Colloquially, these are heterogeneously categorised as \emph{outliers}, but not all outliers are created equal. Some are data flaws (cosmic-ray hits, detector defects), and some are mistakes (data reduction errors, underestimated uncertainties). These are the chaff, which we must recognise only so they do not contaminate the inferred manifold. The other outliers are the fascinating freaks: the rare stars, unlike all others, that can be most informative about astrophysics.

Matrix factorization methods are often used to solve these problems. Principal component analysis \citep[PCA;][]{pearson1901, hotelling1933}, non-negative matrix factorization \citep{lee1999}, heteroskedastic matrix factorization \citep{tsalmantza2012}, and their relatives offer data-driven approaches to simplifying, de-noising, and lowering the dimensionality of complex datasets, without the need for a detailed physical model.
They are generally fast and straightforward to fit with numerical linear algebra or straightforward optimization schemes.
In stellar spectroscopy, these methods are used to find a low-dimensional embedding of the spectra of a population of stars, which can then be used for label transfer \citep{ness2015,casey2016}, classification \citep{refiorentin2007}, or instrument calibration \citep{zhao2021,casey2026}.
In high-contrast imaging, they are used to build a data-driven model of the quasistatic starlight in a sequence of exposures, in order to subtract that signal from each frame and look for faint companions in the stacked residuals \citep{soummer2012, amara2012, fergus2014}.

PCA is the most widely used of these methods, by a long margin. PCA has a long history of success in astronomical spectroscopy. \citet{connolly1995} pioneered the use of PCA for galaxy spectral classification, a technique extended to redshift estimation by \citet{connolly1999}. In stellar spectroscopy, \citet{yip2004} demonstrated the power of PCA-based dimensionality reduction on large spectroscopic surveys such as SDSS. These foundational works established that low-dimensional structure is a powerful organizing principle in spectroscopic datasets. PCA approximates a full-rank rectangular $N \times M$ data matrix ${\bf Y}$ with a matrix ${\bf L}$ of lower rank $K \le {\rm min} \, (N, M)$ by minimizing the sum of squared residuals between ${\bf Y}$ and ${\bf L}$ \citep{eckart1936}.
The $K$ \emph{components} delivered by PCA are orthonormal basis vectors contained in the $K$ rows of some $K \times M$ matrix ${\bf G}$, while the \emph{coefficients} are the $N$ rows of an $N \times K$ matrix ${\bf A}$, such that ${\bf L} = {\bf A}\,{\bf G}$.
PCA simultaneously learns both the basis vectors and the coefficients, and it has the nice property that the basis vectors are orthonormal and ordered by explained variance. Like most matrix factorization methods, PCA requires complete, rectangular data, and treats every data point as equally important. As a consequence, it is extremely sensitive to outliers: a single bad data point in a single row can spoil many or all of the delivered basis vectors.
This is because PCA aims to minimize the unexplained variance, and the empirical variance in a block of data can easily be dominated by a few outliers.

Unrelated to the successes of PCA, weighted least squares (WLS) or $\chi^2$ fitting \citep{legendre1805} has been used extensively, but is similarly sensitive to outliers.
Astronomers have mostly dealt with this using $\sigma$-clipping \citep{hoaglin1983}, which removes data points with large residuals relative to their measurement uncertainties.
This adds \emph{robustness}: the ability of a model to deal with anomalous data points that violate the assumptions of the model.
$\sigma$-clipping is however an ad-hoc fix, and is \emph{conditioned on the assumption} that your current estimate of the model is correct, or at least good enough to identify outliers. For this reason, $\sigma$-clipping is subject to a form of ``mode collapse'' in which large amounts of data are excluded and the inferred model fails to be representative of most of the data.
Fortunately, statisticians developed iteratively reweighted least squares \citep[IRLS;][]{holland1977}, a family of methods that has all the good properties of WLS, while being robust to outliers, and is generally far less prone to mode collapse. IRLS is a workhorse in many domains, including astronomy \citep[e.g.][]{magnier2020}.

\citet{candes2011} originally proposed a \emph{robust PCA}, where the decomposition of ${\bf Y}$ is into both a low-rank component ${\bf L}$ and a sparse component ${\bf S}$, such that ${\bf Y} = {\bf L} + {\bf S}$.
Under some reasonable conditions, \citet{candes2011} showed that this decomposition can be performed exactly.
Since then, the method has been iterated on in many ways, focusing mostly on speed \citep{lin2010}, scalability \citep{rodriguez2013, halko2011}, or including some form of regularisation \citep{guyon2012, wohlberg2012}.
A subset of these many methods have provable convergence \citep{netrapalli2014}, are non-linear \citep{scholkopf1998, wohlberg2012}, or have a probabilistic interpretation \citep{tipping1999}.

The challenge arises when dealing with real data: missing data segments, varying measurement uncertainties across features (heteroskedasticity), and outliers that violate model assumptions. Several methods address the missing-data and heteroskedasticity problem. \citet{budavari2009} developed techniques to handle both missing data and non-uniform weights in a PCA framework. \citet{bailey2012} and \citet{delchambre2015} separately presented methods for PCA with missing data and per-feature uncertainties. These methods successfully handle heteroskedasticity and incompleteness, but lack a principled approach for automatic outlier identification.

\citet{tsalmantza2012} introduced heteroskedastic matrix factorisation (HMF), extending PCA to incorporate both missing data and measurement uncertainties by weighting each point by its inverse uncertainty variance. However, HMF remains sensitive to outliers and anomalies unless they are pre-downweighted, and it does not have a principled way to identify these unknown outliers. At the same time, robust PCA methods (discussed above) handle outliers through sparse decomposition but ignore both missing data and measurement uncertainties.

The method we present in this paper---which we call Robust HMF, or RHMF---is a straightforward generalisation of HMF that addresses both goals at once.
Like HMF, it supports missing data and heteroskedastic measurement uncertainties. Unlike HMF, it adds robustness to outliers through an iteratively reweighted least squares (IRLS) scheme, and automatically downweights entries on both a per-feature (e.g., pixel) and a per-object level.
This automatic downweighting provides a continuous measure of discrepancy (outlier-y-ness, if you will) that can be used to identify anomalous features (pixels) or objects (stars), such as unusual spectra or emission lines.
The algorithm uses a simple alternating linear least squares scheme with provable convergence properties and competitive performance.
We show that the model also has a probabilistic interpretation as either heavy-tailed maximum likelihood estimation, or equivalently as a hierarchical Bayesian model with latent variances.

To our knowledge, no published robust matrix factorization method has all of these properties together: heteroskedasticity, missing data, robustness to outliers, and a probabilistic interpretation.
However, the ingredients we draw on are not individually new. The closest competitor analogue is cellPCA \citep{centofanti2024}, which handles casewise outliers, cellwise outliers, and missing entries in a single IRLS objective, but assumes homoskedasticity.
Heavy-tailed and IRLS-based approaches to robust subspace recovery have a long history more broadly \citep[for reviews see][]{bouwmans2015,polyak2017}. Unlike \citet{candes2011} and its descendants \citep[e.g.,][]{zhou2010,yin2019}, we do not explicitly learn a sparse outlier matrix ${\bf S}$, and instead rely on the inferred robust weights to quantify outlier-y-ness. 
The loss family we use is also a direct descendant of Truncated Cauchy NMF \citep{guan2019}, since fixing $\nu=1$ in our Student-t likelihood recovers a Cauchy loss.
What RHMF adds to this body of work is the combination, at survey scale, of: a probabilistic / hierarchical-Bayes formulation that exposes the latent variances; per-pixel heteroskedasticity carried through the model from the start, rather than added as a post-hoc reweighting of an outlier-only loss; coherent variational down-weighting at both the pixel and the object level inside a single objective; and a \texttt{JAX} implementation that scales to spectroscopic-survey-size data. Our novelty claim is therefore the package of these properties at scale, not any one of them in isolation.

In Section~\ref{sec:methods} we present the model, the algorithm by which it is fit, and discuss using it for outlier identification. 
In Section~\ref{sec:experiments} we demonstrate the approach on two example problems: a toy dataset, and with Gaia RVS spectra. 
The toy dataset is provided for illustration and validation of the method, and the Gaia RVS spectra demonstrate outlier recognition and triage. 
In Section~\ref{sec:discussion} we discuss the strengths and shortcomings of the model, potential applications and extensions, and the relationship to other methods in the literature. 

\section{Method} \label{sec:methods}

\subsection{Model setup and assumptions} \label{sec:assumptions}

Let $N \in \mathbb{Z}^+$ be the number of objects (in our case, spectra) in the dataset, and $M\in\mathbb{Z}^+$ be the number of features (pixels) for each object.
We then label the value in feature $j$ of object $i$ as $y_{ij}$, and we can also write the data as an $N \times M$ matrix ${\bf Y}$ with entries $y_{ij}$.
We assume that the investigator has measurement uncertainties $\sigma_{ij}$ corresponding to each feature, and also that these uncertainties have the same units as $y_{ij}$.
In practice we will often refer to ``weights'' instead, which are simply inverse variances $w_{ij} = \sigma_{ij}^{-2}$.

The model for the data is
\begin{align}
    y_{ij} &= \sum_{k=1}^K a_{ik}\,g_{kj} + \rm{outliers} + \rm{noise}, \\
    \rm{or \,\, equivalently} \quad {\bf Y} &= {\bf A}\,{\bf G} + \rm{outliers} + \rm{noise}, \label{eq:model}
\end{align}
which is a low dimensional linear embedding of rank $K\in\mathbb{Z}^+$.
$a_{ik}$ are the entries of an $N \times K$ matrix ${\bf A}$ where each of the $N$ rows contains $K$ \emph{coefficients}, and $g_{kj}$ are the entries of a $K \times M$ matrix ${\bf G}$ where each of the $K$ rows contains \emph{basis vectors} of length $M$.
The outliers appear explicitly in the above, but we do not model them as an explicit component unlike many other robust PCA methods as already discussed.
The choice of the rank $K$, or the number of basis vectors to include, is a hyperparameter to be set by the user.
Because of the objective function we will adopt, and unlike with PCA, the rank $K$ cannot be chosen \emph{after} fitting.

We make the following assumptions:
\begin{enumerate}
    \item \textbf{Heteroskedasticity}: Each feature $j$ of each object $i$ has a known measurement uncertainty $\sigma_{ij}$ that is the root-variance of a zero-mean effective or expected Gaussian noise distribution. Each of these uncertainties $\sigma_{ij}$ can be different for each different $i,j$. In what follows we will often speak in terms of inverse-variance weights $w_{ij}=\sigma_{ij}^{-2}$; data with large uncertainties have low weights.
    \item \textbf{Unreliable measurement uncertainties}: The measurement uncertainties $\sigma_{ij}$ are known but they may not all be \emph{accurate} in that some $y_{ij}$ may be outliers or have underestimated $\sigma_{ij}$.
    \item \textbf{Missing data}: Missing values at particular locations $i, j$ can be handled by setting weights $w_{ij}$ to zero or (equivalently) by setting the corresponding measurement uncertainties $\sigma_{ij}$ to infinity (or arbitrarily large).
    \item \textbf{Rectangular data}: At each fixed feature index $j$, the values across all objects correspond to the same $j$-th feature.
    \item \textbf{Basis orthogonality}: The inferred basis vectors, and so the rows of ${\bf G}$, will be made strictly orthogonal and ordered by explained variance, in analogy to principal component analysis.
      \item \textbf{Low rank}: The model is strictly low rank, with $1 \le K < \min(N, M)$. When $K = \min(N, M)$, the product ${\bf A}\,{\bf G}$ can represent any $N \times M$ matrix and the model perfectly reproduces ${\bf Y}$ with zero residuals everywhere. In that case all robust weights collapse to unity, no outliers are identified, and no noise is separated from signal---the model simply memorizes the data. Strict low rank forces the model to choose which structure to retain and which to treat as residual, and it is this compression that makes the robustness and outlier detection meaningful.
    \item \textbf{Outlier sparsity}: The outliers in the data are rare; the vast majority of the nonzero-weight pixels are \emph{not} outliers.
    \item \textbf{Outlier uniqueness}: The outliers do not themselves show low-rank structure; that is, the low-rank structure is only in the \emph{inlier} and not the \emph{outlier} part of the data set.
\end{enumerate}

In the end, although we are thinking about having outlier features, we will in fact inflate the uncertainties (lower the weights) on such features such that they are no longer outliers.
Thus the model does not explicitly classify features into binary outlier and non-outlier classes.
This distinguishes our approach from standard low-rank-plus-sparse approaches \citep{candes2011}.
Related to this, our method is not convex, so there is an additional implicit assumption that the optimization scheme, including its initialization, finds a good local solution.
This is generally easy when the outliers are rare and the heteroskedasticity is not extreme.

We emphasize that ``outliers'' in the RHMF framework refers to deviations from the assumed low-rank structure, \emph{not} to objects or features that belong to a different class or category. This distinction is crucial for realistic data. For example, consider spectroscopic data from a sample containing four stellar types with population counts of 100, 100, 50, and 10. The rare type with 10 objects would \emph{not} be classified as outliers by RHMF, even if spectrally distinct from the other three types, provided that those 10 spectra share a coherent low-rank structure among themselves (i.e., they follow the low-rank subspace). Instead, the model would allocate additional basis vectors to capture the shared spectral variations of the rare type. Conversely, a single spectrum that deviates significantly from the common low-rank structure of its class would be downweighted, even if it belongs to a populous class. Moreover, rarity does not imply outlier-ness: extremely metal-poor stars are rare, but they would not be identified as outliers if their spectra can be described by the existing low-rank basis. Outliers are those that cannot be described well by the low-rank structure.

When data contains multiple well-separated populations with balanced or substantial counts, the model behavior depends on the choice of rank $K$. Lower $K$ will prioritize the most common population and treat the others as collectively outlier-like; higher $K$ will recover the distinct low-rank structure within each population, at the cost of reduced robustness to true anomalies. There is no principled statistical way to distinguish between these two interpretations from the data alone---it requires domain knowledge about which structure is ``interesting'' and which is ``noise''. We recommend users who suspect multimodal data inspect both the basis vectors and the structure of the residuals: coherent residual patterns shared across many objects suggest additional low-rank structure that higher $K$ could capture. If the investigator wishes to treat separate populations symmetrically and detect anomalies within each, increasing $K$ to accommodate all populations is a reasonable approach, with the understanding that the robustness to true outliers may diminish slightly.

\subsection{Robustness}

Models in this matrix-factorization family (e.g., PCA, where all $w_{ij} = 1$, and HMF), are usually fit by minimizing the $\chi^2$ metric, or equivalently maximizing a Gaussian likelihood
\begin{align}
    \hat{\bf A}, \hat{\bf G} &= \argmin_{{\bf A},{\bf G}} \left[\sum_{ij} w_{ij}\left(y_{ij} - \sum_k a_{ik} g_{kj} \right)^2\right], \\
    {\rm or \,\, equivalently} \quad y_{ij} &\sim {\rm Normal} \left( \sum_k a_{ik} g_{kj}, \sigma^2_{ij}\right), \label{eq:gaussian_like}
\end{align}
where the hat notation denotes the best-fitting values of ${\bf A}$ and ${\bf G}$, and Equation~\ref{eq:gaussian_like} denotes that each $y_{ij}$ is \emph{drawn from} a Normal distribution with mean $a_{ik} g_{kj}$ and variance $\sigma_{ij}^2$.
These two lines are equivalent; the top line minimizes the negative log likelihood defined implicitly by the second line.
This setup induces a quadratic penalty in the residuals $r_{ij} = y_{ij} - \sum_k a_{ik}g_{kj}$, which causes outliers to have a large influence on the fit.

We introduce robustness to the model by replacing the likelihood with a heavy-tailed distribution.
Here, we choose the Student-t, but our method is generalizable to any heavy-tailed distribution.
The cost function (negative log likelihood) and likelihood therefore becomes
\begin{align}
    \hat{\bf A}, \hat{\bf G} &= \argmin_{\bf A, G} \left[ \sum_{ij} \log\left(1 + \frac{w_{ij} r_{ij}^2}{\nu s^2}\right) \right], \label{eq:robust_opt} \\
    {\rm or \,\, equivalently} \quad y_{ij} &\sim {\rm StudentT}_\nu \left( \sum_k a_{ik} g_{kj}, s^2 \sigma_{ij}^2 \right) \label{eq:t_like}
\end{align}
where the number of degrees of freedom $\nu\in\mathbb{Z}^+$ and the scale $s\in\mathbb{R}^+$ are parameters that control the heaviness of the distribution's tails.
In the limit $\nu \rightarrow \infty$ with $s=1$ this converges to the Normal likelihood.
This modification to the cost function results in a quadratic penalty for small residuals, and a logarithmic penalty for large residuals, reducing the influence of outliers on the fit.

This setup can be equivalently viewed as a hierarchical model with latent, unknown variances $\tau_{ij}^2$
\begin{align}
    y_{ij} &\sim {\rm Normal} \left( \sum_k a_{ik} g_{kj}, \tau_{ij}^2 \right), \label{eq:hierachical_like} \\
    \tau_{ij}^2 &\sim {\rm InverseGamma} \left( \frac{\nu}{2}, \frac{\nu}{2} s^2 \sigma_{ij}^2 \right) \label{eq:hierarchical_prior}
\end{align}
where the inverse Gamma distribution is a strictly positive distribution with shape $\frac{\nu}{2}$ and scale $\frac{\nu}{2} s^2 \sigma_{ij}^2$ parameters.
The equivalence between the Student-t likelihood and the hierarchical view is due to the existence of a closed-form marginal likelihood for this choice of prior; marginalizing $\tau_{ij}$ from Eq.~\eqref{eq:hierachical_like} with Eq.~\eqref{eq:hierarchical_prior} yields Eq.~\eqref{eq:t_like} \citep[e.g.][]{lange1989}.

For the applications we present here, $s$ and $\nu$ are not uniquely identifiable.
This is because the likelihood depends only on the product $Q^2=\nu s^2$ (Eq.~\ref{eq:t_like}), so our Student-t maximum likelihood estimate (MLE) cannot distinguish them. 
We therefore fix $\nu=1$ (which reduces to the Cauchy distribution) without loss of generality.

After the number of basis functions $K$, $Q$ now represents  the second hyperparameter of the model: a soft outlier threshold that controls where the transition between the quadratic and logarithmic penalties occurs.
With $Q=1$ the transition occurs at about the same point as the measurement uncertainties, ${Q}>1$ means that the transition occurs at larger residuals than the measurement uncertainties, and ${Q}<1$ means that the transition occurs at smaller residuals than the measurement uncertainties. 

\subsection{Fitting Algorithm} \label{sec:fitting}

While in principle it's possible to optimize Eq.~\eqref{eq:robust_opt} for all $a_{ik}$ and $g_{kj}$ simultaneously, we invoke a majorization-maximization scheme that implicitly minimizes the objective by iterating between re-weighting and least-squares solves.
This approach is exactly the typical IRLS method for robust regression, but applied to HMF instead of linear regression.
We prove that the algorithm below optimises the likelihood defined in Eq.~\eqref{eq:t_like} in Appendix~\ref{sec:proof}.

While the algorithm we present is guaranteed to converge to a local optimum, it is not guaranteed to converge to the global optimum, and so it is pertinent to provide a good initialization.
The most straightforward approach is the singular-value decomposition
\begin{align}
    \left[{\bf Y}\right]_{ij} &= y_{ij}, \\
    {\bf U} {\bf S} {\bf V}^\top &= \svd{\left({\bf Y}\right)}, \\
    a_{ik} &\leftarrow \left[{\bf S}\right]_{kk}^{1/2} [{\bf U}]_{ik}, \\
    g_{kj} &\leftarrow \left[{\bf S}\right]_{kk}^{1/2} [{\bf V}^\top]_{kj},
\end{align}
where ${\bf U}$ and ${\bf V}^\top$ are unitary matrices, and {\bf S} is a matrix with diagonal entries equal to the singular values of ${\bf Y}$ in non-decreasing order, and zeros elsewhere.
This initialization is simply the first $K$ components of the PCA solution.
In the case that the dataset is very large, it may be preferable to use a faster, approximate method such as the randomized SVD \citep{halko2011}.

Unlike standard PCA, RHMF does not require mean subtraction of the data matrix ${\bf Y}$ before computing the SVD initialization. The first basis vector (first row of ${\bf G}$) is free to capture any global offset or mean level in the data, just as any other basis vector. This is a natural consequence of the flexible linear model in Eq.~\eqref{eq:model}: the model can allocate coefficients to represent the mean without explicit pre-processing. In practice, this means RHMF can be applied directly to data as observed, making it more convenient than PCA when the mean is scientifically meaningful or when multiple datasets with different mean levels need to be compared within a single low-rank representation.

Iteratively updating the parameters consists of four steps that cycle until convergence. The first is the \textbf{w-step}, which  updates the data weights to downweight outliers, given the current estimate of ${\bf A}$ and ${\bf G}$:
\begin{align}
    r_{ij} &= {y_{ij} - \sum_k a_{ik} g_{kj}}\\
    w^{\rm robust}_{ij} & = \frac{Q^2}{Q^2 + {w_{ij}^{\rm data}}\,r_{ij}^2}, \label{eq:w_robust} \\
    w^{\rm (new)}_{ij} &\leftarrow w^{\rm data}_{ij}\, w^{\rm robust}_{ij},
\end{align}
where we note that $w^{\rm robust}_{ij} \in (0, 1]$ provides a measure of outlier-y-ness on the $j$th feature in the $i$th object, and that $w_{ij}^{\rm (new)} = \tau_{ij}^{-2}$.
This rule also respects data weights of zero, and shows clearly how $Q$ acts to control how aggressively large residuals cause downweighting.

The \textbf{a-step} finds the best-fit values for the coefficients $a_{ik}$ given the current estimate of the basis vectors $g_{kj}$:
\begin{align}
    [{\bf X}_i]_{kk'} &= \sum_{j=1}^M g_{kj} w_{ij}^{\rm (new)} g_{k'j}, \\
    [{\bf b}_i]_k &= \sum_{j=1}^M g_{kj} w_{ij}^{\rm (new)} y_{ij}, \\
    \boldsymbol{\alpha}_i &= \solve{\left( {\bf X}_i, {\bf b}_i \right)}, \\
   a_{ik}^{\rm (new)} &\leftarrow [{\boldsymbol{\alpha}}_i]_k
\end{align}
where the operator $\solve({\bf X}, {\bf b})$ returns ${\bf X}^{-1}\,{\bf b}$.
This is just the WLS solution for the rows of ${\bf Y}$ given fixed ${\bf G}$.

Next, the \textbf{g-step} finds the best-fit basis vectors $g_{kj}$ given the current estimate of the coefficients $a_{ik}$:
\begin{align}
    [{\bf X}_j]_{kk'} &= \sum_{i=1}^N a^{\rm (new)}_{ik} w_{ij}^{\rm (new)} a^{\rm (new)}_{ik'}, \\
    [{\bf b}_j]_k &= \sum_{i=1}^N a^{\rm (new)}_{ik} w_{ij}^{\rm (new)} y_{ij}, \\
    \boldsymbol{\gamma}_j &= \solve{\left( {\bf X}_j, {\bf b}_j \right)}, \\
   g_{kj}^{\rm (new)} &\leftarrow [\boldsymbol{\gamma}_j]_k
\end{align}
which is just the WLS solution for the columns of ${\bf Y}$ given fixed ${\bf A}$.

The \textbf{rotation-step} suppresses the huge set of degeneracies in the model by enforcing a standard orientation in either the data or the model space. Here, we will require that the basis vectors be orthonormal:
\begin{align}
    \boldsymbol{\lambda}, {\bf V} &= \eig{\left( {\bf G} {\bf G}^\top + \epsilon {\bf \, I}\right)},\\
    {\bf A}^{\rm (new)} &\leftarrow {\bf A} {\bf V} \, {\rm diag} \left(\boldsymbol{\lambda}^{1/2} \right) {\bf V}^\top, \\
    {\bf G}^{\rm (new)} &\leftarrow {\bf V} \, {\rm diag} \left(\boldsymbol{\lambda}^{-1/2}\right)  {\bf V}^\top {\bf G},
\end{align}
where the operator ${\rm eig}(\cdot)$ returns a $K$-vector and a $K\times K$ matrix, containing eigenvalues and eigenvectors respectively.
$\epsilon$ is a small value (we use $10^{-6}$) added to the diagonal of ${\bf G} {\bf G}^\top$ to avoid numerical issues that can sometimes occur near the start of training, due to $\mathbf{G}\mathbf{G}^\top$ having near-zero eigenvalues\footnote{For the same reason, also explicitly floor all eigenvalues $[\boldsymbol{\lambda}]_k$ to $\epsilon$ after the eigendecomposition.}.
Note that here the eigendecomposition is guaranteed to be performed on a real and symmetric matrix, and so allows for slightly faster numerical routines than the general case.

We assess convergence every few cycles by a dimensionless estimate of the size of the g-step adjustment.
The outputs of the procedure are the full matrices ${\bf A}$ and ${\bf G}$.
The robust weights $w^{\rm robust}_{ij}$ for any data point are also calculable by Eq.~\eqref{eq:w_robust}.
Inference on new or held-out data is performed simply by reusing ${\bf G}$, and performing the a- and w-steps (such that ${\bf G}$ is never modified) until convergence.
When iterating on new or held-out data, we assess convergence by a dimensionless estimate of the size of the a-step adjustment instead.
The asymptotic complexity of this algorithm per cycle, assuming that $K \ll N, M$, is $\mathcal{O}(NMK^2)$, which is linear in both $N$ and $M$.

\subsection{Validation and hyperparameter choice} \label{sec:cv}

As part of fitting the model, the investigator is required to choose the hyperparameters $Q\in \mathbb{R}^+$ and $K\in \mathbb{Z}^+$.
Depending on the application, it may be sufficient to set these to reasonable values and move on, but in some cases it may require careful thought.
We discuss when hyperparameter choice matters further in Section~\ref{sec:discussion}.
Assuming that it does matter, we advocate for a held-out data cross-validation approach which is both straightforward and prior-independent.
The data should be partitioned randomly into a training set, and the model should be fit on the training set only.
This should be repeated over a grid of $Q$ and $K$ values, and the best-fitting model for each pair of hyperparameters should be used to predict the test set data.

We propose evaluating the model by measuring how closely the weighted residuals on the held-out data resemble a unit normal distribution.
Under the model, the standardised residuals $r_{ij}\sqrt{w^{\rm data}_{ij}}$ for inlier data points are drawn from $\mathcal{N}(0,1)$ by construction.
The inferred robust weights $w^{\rm robust}_{ij}$ further shrink outlier residuals toward zero, so the combined quantity
\begin{equation}
    z_{ij} = r_{ij}\sqrt{w^{\rm data}_{ij}\,w^{\rm robust}_{ij}}
\end{equation}
should follow $\mathcal{N}(0,1)$ across the whole held-out population when the model is well calibrated and the outliers are correctly identified.
We therefore define the cross-validation score as the Kullback--Leibler (KL) divergence \citep{kullback1951} from the empirical distribution of $z_{ij}$ to $\mathcal{N}(0,1)$.
Approximating the empirical distribution as Gaussian with sample mean $\hat{\mu}_z$ and sample standard deviation $\hat{\sigma}_z$, the closed-form expression is
\begin{equation}
    \score(Q,K) = \mathrm{KL}\!\left(\mathcal{N}(\hat{\mu}_z,\,\hat{\sigma}_z^2)\,\Big\|\,\mathcal{N}(0,1)\right) = -\log\hat{\sigma}_z + \frac{\hat{\sigma}_z^2 + \hat{\mu}_z^2}{2} - \frac{1}{2}, \label{eq:score}
\end{equation}
which equals zero when $\hat{\mu}_z = 0$ and $\hat{\sigma}_z = 1$, and is strictly positive otherwise.
This score has well-defined behaviour as $K$ grows and can be directly compared across $(Q,K)$ pairs: models that correctly capture the shared structure and correctly identify outliers will produce $z_{ij}$ values that more closely resemble unit normals, and hence smaller scores.

\subsection{Feature- and object-level outlier identification} \label{sec:outlier_id}

The inferred  \emph{robust} weights $w^{\rm robust}_{ij}$ take a value between 0 and 1, with $\rightarrow0$ corresponding to complete downweighting (an outlier) and $\rightarrow1$ corresponding to no downweighting (an inlier).
This property means that the individual values of the weights can effectively give a continuous, per-feature-per-object measure of the degree to which a data point is an outlier.
One can do this by assembling statistics or thresholds relative to the distribution of all $w^{\rm robust}_{ij}$.
Furthermore, per-object or per-feature measures of outlier-y-ness can be assembled from some summary statistic over either the $j$ or $i$ index respectively.
A simple example is to take the per-object median
\begin{align}
    w^{\rm object}_i &= \operatorname*{median}_{j} \left( w_{ij}^{\rm robust} \right), \label{eq:w_spec}
\end{align}
and then to look at the distribution of $w^{\rm object}_i$ across the data set to see the degree to which any one object deviates from the population.

One is free to choose whatever metric over $w^{\rm robust}_{ij}$ best suits a particular dataset.
For example, in stellar spectra, the median weight per-object could be a bad measure, in that the spectra of interest could be mostly typical, but with one unusual line.
The minimum or the 0.01 quantile weight per object, might be better choices in that case.
One could also look for contiguous stretches of pixels (features) with an average weight below a threshold.
For example, if a goal is to find objects that are unusual in particular features (e.g., specific absorption lines in a star), then the measure could be the median weight per-object over some narrow wavelength range.
We will demonstrate using the median per-object weight in the toy example (Section~\ref{sec:toy}), and using the 0.01 quantile in the \textsl{Gaia} examples (Sections~\ref{sec:gaia} and \ref{sec:gaia_full}).

\subsection{Implementation}

We implemented the RHMF algorithm in an open-source Python package called \texttt{Robusta-HMF} \citep{hilder2026}.
\texttt{Robusta-HMF} is designed both for those who want an out-of-the-box solution for RHMF (and HMF), and also for those who want to extend/modify the algorithm for their own purposes.
The code is written in \texttt{JAX} \citep{jax2018github}, which provides just-in-time compilation, GPU and TPU support, and automatic differentiation.
The API is designed to be easy to use, with a \texttt{scikit-learn}-esque \citep{scikit-learn} interface for fitting the model and making predictions.

The package additionally supports custom initialization and convergence criteria, and is easily extensible to other robust losses and likelihood functions, as well as adding regularization or other constraints on the model.
It also supports direct optimization of the likelihood with gradient-based methods as an alternative to the IRLS scheme, which may be preferable in some cases (although beyond the scope of this paper).

\texttt{Robusta-HMF} is available for installation from the Python Packaging Index, e.g. \texttt{pip install robusta-hmf} or \texttt{uv add robusta-hmf}.
It is also available at \url{https://github.com/TomHilder/robusta-hmf}, and the code to reproduce both experiments in the following sections is available in the same repository.

\section{Experiments} \label{sec:experiments}

\subsection{Toy Spectra} \label{sec:toy}

We generated 8,000 synthetic spectra (objects) with 1,200 pixels (features) on the same wavelength grid, from a linear sum of a few polynomials and a low- and fixed-frequency sinusoid to represent continuum, and absorption lines present in all the spectra.
The true underlying model is therefore linear with rank 5.
We then corrupted the dataset in following ways:
\begin{enumerate}
    \item We added Gaussian noise with a standard deviation proportional to a random factor across spectra, a systematic factor across wavelength, and an additional random factor per pixel. We take these standard deviations as known measurement uncertainties for the analysis.
    \item One contiguous missing data segment was injected in half of all spectra, at a random wavelength location, with a length between 50 and 200 pixels.
    \item 40 of the 8,000 spectra were constructed to be \emph{outlier spectra}: with a high frequency sinusoid and a randomly drawn frequency and amplitude in some range. This set contains little shared low-rank structure between them, and so should be interpreted as representing outlier spectra of different kinds, rather than some second class of spectra with similar properties.
    The outlier spectra may also contain missing data segments, and the other types of outliers described below.
    \item We injected \emph{outlier columns}, which are fixed pixel indices at which 30\% of all spectra contain a random large (or very negative) value, intended to represent some systematic reduction, calibration, or instrument issue that affects a particular wavelength across many spectra.
    \item We injected individual \emph{outlier pixels} at random locations, consisting of a fixed 0.4\% of all pixels in the data. We did not set the corresponding measurement weights to zero as these are intended to represent \emph{unknown} bad pixels.
    \item In 10 non-outlier spectra (hereafter \emph{normal spectra}) we injected 3 absorption lines at random wavelength locations with random amplitudes. We refer to these as \emph{outlier lines}.
\end{enumerate}
The exact details of the basis function functional form, the random sampling used for the coefficients, and the random sampling used to generate the noise, missing data, and each type of outlier, is provided in Appendix~\ref{sec:toy_model_details}.

We fit the model with 35 different combinations of the hyperparameters $Q$ and $K$ to the same random subset of 4,000 spectra, with the other 4,000 held out for validation.
The values of $Q$ were chosen to span a range from very aggressive downweighting of outliers ($Q=0.5$) to very little downweighting at all ($Q=10$), and the values of $K$ were chosen to span a range from underfitting ($K=3$) to overfitting ($K=7$) the true underlying model.
The values we tested were $Q\in\left\{ 0.5, 1, 2, 3, 4, 5, 10\right\}$ and $K\in\left\{ 3, 4, 5, 6, 7\right\}$.
For each fit, we initialized with an SVD, and iterated the algorithm until the maximum fractional change in any entry of ${\bf G}$ was less than $10^{-2}$, checking for convergence every 10 cycles.

We also fit a standard PCA model, and a robust PCA (RPCA) model using the method of \citet{candes2011} as implemented in the \texttt{robust-pca} Python package \citep{ganguli2026}.
For both of these we used the same training set as in the RHMF case, and stuck to the true underlying model rank of $K=5$ as a best-case scenario for these methods.
Since these methods do not support providing data weights nor missing data, we set the missing data points to zero and ignored the measurement uncertainties.
Given that the data points in the spectra are all near 1, it might be more realistic for this case to set the missing data points to 1, or to the mean of the dataset, or similar.
However in full generality the investigator may not have a good way to set the missing data, and so we chose to set them to zero for illustrative purposes.
Indeed, if the missing data segment contained an absorption line (which some of them do), setting the missing data to 1 or the mean would be a similarly bad choice.

\begin{figure}
    \centering
    \includegraphics[width=0.98\linewidth]{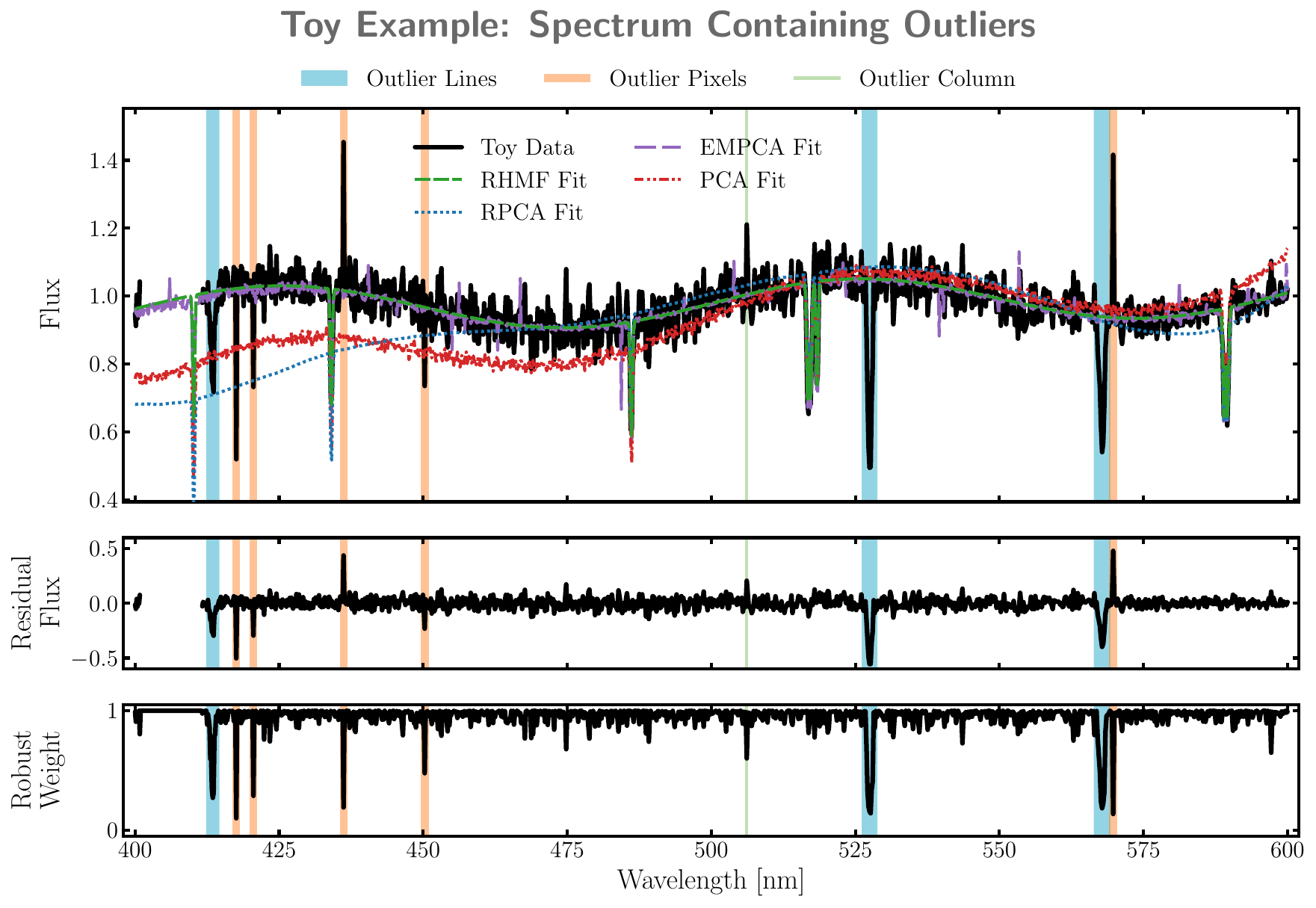}
    \caption{Randomly selected noisy toy spectrum containing outlier absorption lines, random outlier bad pixels, a pixel bad across many spectra, and a missing data segment. Top panel shows the data compared with the best-fit PCA (red, dashed-dotted), RPCA (blue, dotted), EMPCA (purple, long-dashed), and RHMF (green, dashed) models. Middle panel shows the RHMF model-subtracted residuals, demonstrating that the model has ignored the outlier pixels and lines, making them easy to identify by eye. Bottom panel shows the inferred robust weights, showing that the outlier pixels and lines have been down-weighted by the model, also useful for identification. The PCA, RPCA, EMPCA, and RHMF models are all fit with $K=5$ basis vectors, and the RHMF model is fit with $Q=5$. EMPCA, like RHMF, is given the inverse-variance weights.}
    \label{fig:toy_residuals}
\end{figure}

Figure~\ref{fig:toy_residuals} shows a randomly selected spectrum from amongst the 10 normal spectra containing outlier lines.
It also contains 5 outlier pixels, an outlier column, and a missing data segment that covers a region with an absorption line.
The true, known locations of the outliers are indicated by the vertical bands in all panels.
The top panel compares the data, the best-fit RHMF model (with $Q=5$ and $K=5$ as selected by the validation results, see below), the best-fit PCA model, the best-fit RPCA model, and the best-fit EMPCA model.
The PCA model (red) is very noisy despite the underlying model being truly linear, low-rank, and simple, because the basis vectors are ``used up'' trying to explain a lot of empirical variance caused by the presence of outliers.
The PCA model thus fails to capture the underlying structure.
It is also unable to handle the missing data, causing the model to completely fail to reproduce the data near that region.
The EMPCA model (purple) is given the same inverse-variance weights RHMF uses, both when fitting its basis and when projecting this spectrum onto it, so it accounts for the heteroskedastic noise and interpolates across the missing data. Because EMPCA has no outlier mechanism, it is still pulled toward the outlier pixels and lines, and its fit deviates from the underlying structure at those locations.
The RPCA model (blue) fares better at capturing the underlying structure due to the robustness.
However, it is still negatively impacted by the heteroskedasticity, and it deviates from the data severely in the missing data region and mildly elsewhere.
The RHMF model (green) captures the underlying structure very well, as it is able to ignore the outliers, handle the heteroskedasticity, and handle the missing data.
The correct low-rank structure is captured, the model is smooth and continuous, the lines fit the data well, and the model interpolates well across the missing data region.

The middle panel of Fig~\ref{fig:toy_residuals} shows the residuals after subtracting the best-fit RHMF model from the data.
At all regions where outliers are present, the residuals are large, as is intended since the model deliberately does not capture them.
Elsewhere the residuals are small and contain only noise.
The outlier lines are easy to identify by eye in the residuals.
The bottom panel of Fig~\ref{fig:toy_residuals} shows the inferred robust weights $w_{ij}^{\rm robust}$.
These weights exhibit random fluctuations near 1 across most of the spectrum, but drop to values of $\lesssim 0.5$ in outlier locations, with the outlier lines dropping even to $\lesssim 0.2$.

\begin{figure}
    \centering
    \includegraphics[width=0.88\linewidth]{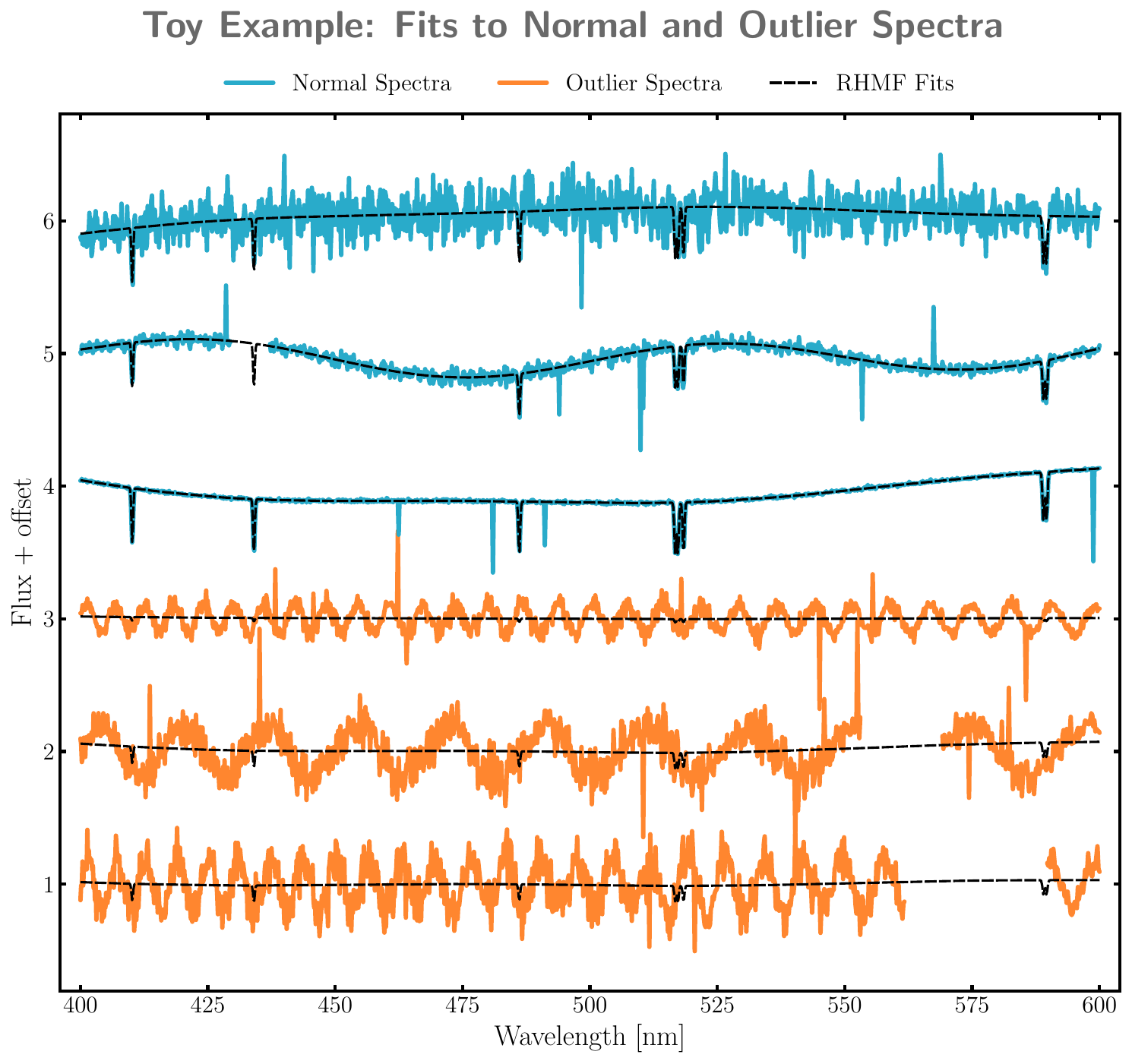}
    \caption{Noisy toy spectra randomly selected from across the training and test sets, with the best-fit RHMF model predictions (black, dashed) overlaid. The top 3 are ``normal'' spectra (blue), while the bottom 3 are outlier spectra (orange). In the normal spectra, the model picks up on the common structure and ignores pixel-level outliers, while in the outlier spectra the model does not attempt to fit the high-frequency sinusoidal structure at all, which is the intent. The shown model has $K=5$ and $Q=5$.}
    \label{fig:toy_spectra}
\end{figure}

Figure~\ref{fig:toy_spectra} shows 6 spectra selected randomly from the whole dataset.
The top three are normal spectra, the bottom three are outlier spectra, and the RHMF best-fit to each is overlaid.
For the normal spectra (blue), the model accurately captures the underlying model, despite the outlier contamination, missing data, and heteroskedastic noise.
For the outlier spectra (orange), the model completely ignores their structure.
This shows that the downweighting mechanism works on both a local per-data-point level, but also across whole spectra that have little shared underlying structure with the rest of the dataset.

\begin{figure}
    \centering
    \includegraphics[width=0.55\linewidth]{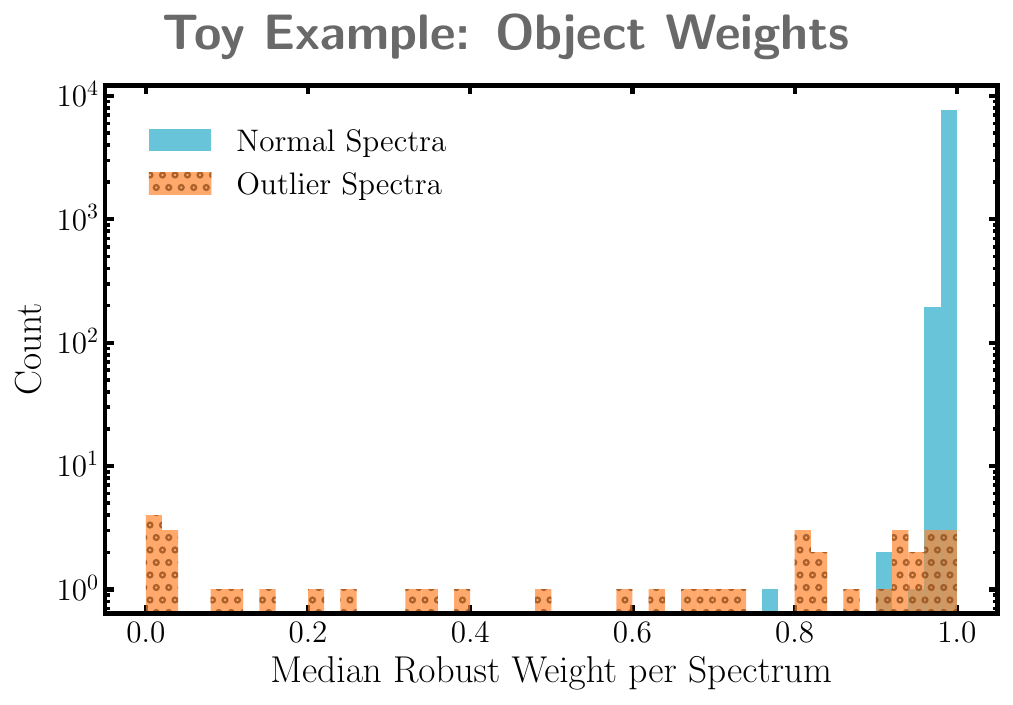}
    \caption{Distribution of the object-level weights $w_i^{\rm object}$ for the toy data, coloured by whether the spectrum is truly outlier (orange, hatched) or not (blue). The model clearly distinguishes the outlier spectra, which tend to have lower weights than normal spectra. The shown model has $K=5$ and $Q=5$.}
    \label{fig:toy_weights}
\end{figure}

To demonstrate outlier spectrum identification, we show the per-object median robust weights $w_i^{\rm object}$ calculated with Eq.~\eqref{eq:w_spec} for the dataset.
These are split into the two known true labels of either ``normal'' (blue) or ``outlier'' (orange).
This includes both the training and test sets.
While these two populations are not perfectly segregated, we see that \emph{all} the normal spectra retain median weights near unity.
In contrast, the outlier spectra have median weights mostly $\leq 0.8$, with only a few in the $> 0.9$ region.

\begin{figure}
    \centering
    \includegraphics[width=0.84\linewidth]{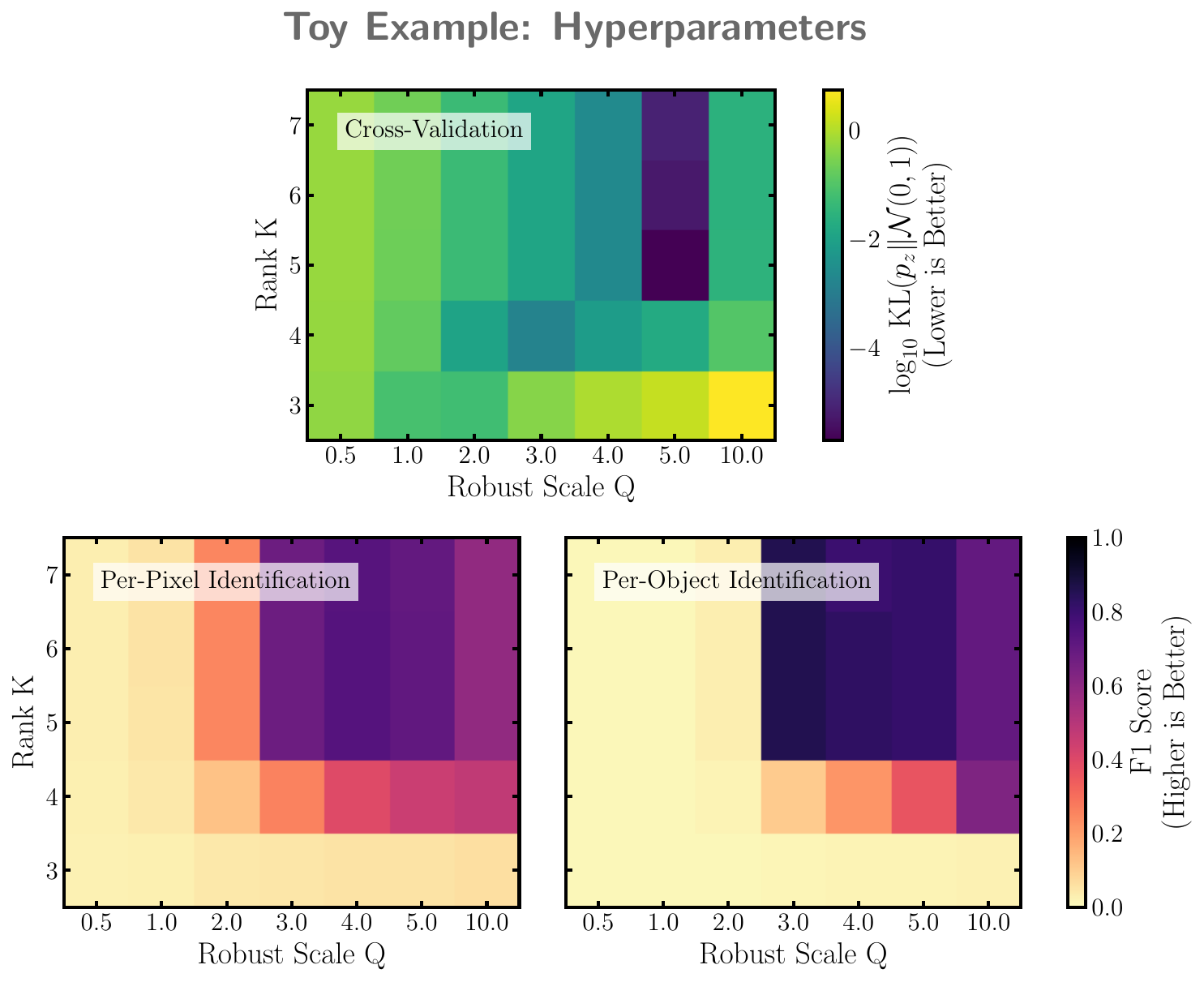}
    \caption{Top: Cross-validation scores (KL divergence from $\mathcal{N}(0,1)$, Eq.~\eqref{eq:score}) for the grid of $Q$ and $K$ values. The score is minimized at $K=5$ and $Q=5$, which is the model shown in the other plots. Bottom left: F1 score for outlier identification on a per-object-per-pixel basis, based on a threshold of $0.5$ on the robust weights $w_{ij}^{\rm robust}$. Bottom right: F1 score for outlier identification on a per-object basis, based on a threshold of $0.9$ on the object-level weights $w_i^{\rm object}$. The F1 scores are calculated with respect to the known true outlier labels, and they demonstrate that both the best identification is obtained near the best validation score, and that picking $K$ too large does not dramatically degrade the outlier identification}
    \label{fig:cv}
\end{figure}

\begin{figure}
    \centering
    \includegraphics[width=\linewidth]{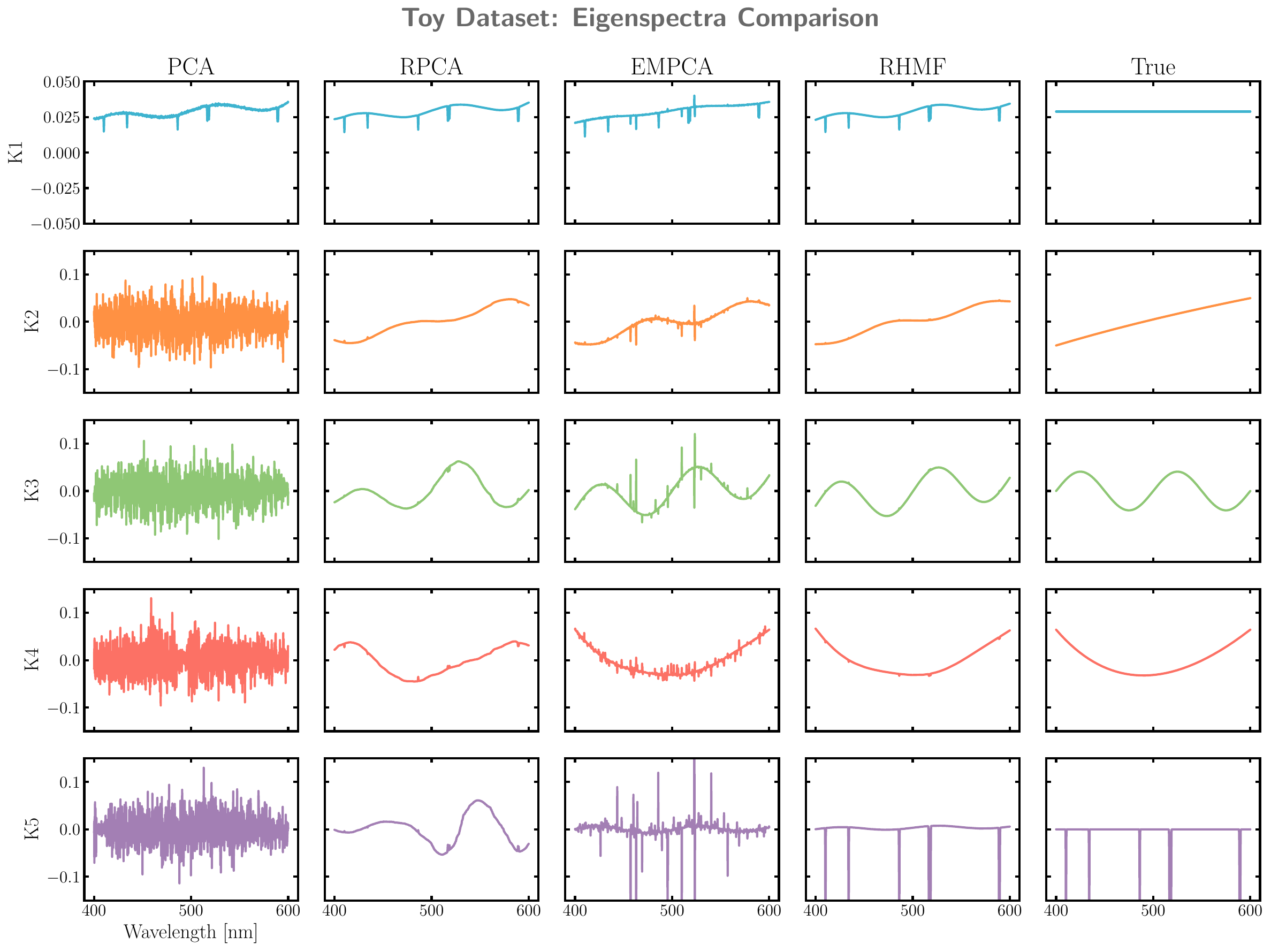}
    \caption{Basis vector comparison across PCA, RPCA, EMPCA, RHMF, and the ground truth for the toy dataset. Each row shows one of the first 5 basis functions (K1--K5), with each column representing a different method; all methods are fit on the same training split, and EMPCA additionally receives the inverse-variance weights. Because any matrix factorization is identified only up to an invertible $K \times K$ transformation, the RHMF basis is rotated to the principal axes of its coefficients (ordered by coefficient power). This is the same canonical frame that the SVD provides for the other methods. The true basis is ordered by its data-power contribution. Component signs are aligned to the truth for all methods. In this common frame, PCA is noise-dominated beyond the first component; EMPCA recovers the smooth structure but is contaminated by outlier residuals (weights encode known noise, not outliers) and fails to isolate the absorption-line template; RPCA is smooth but misses the line template entirely; RHMF recovers all five true components.}
    \label{fig:eigenspectra}
\end{figure}

\begin{figure}
    \centering
    \includegraphics[width=\linewidth]{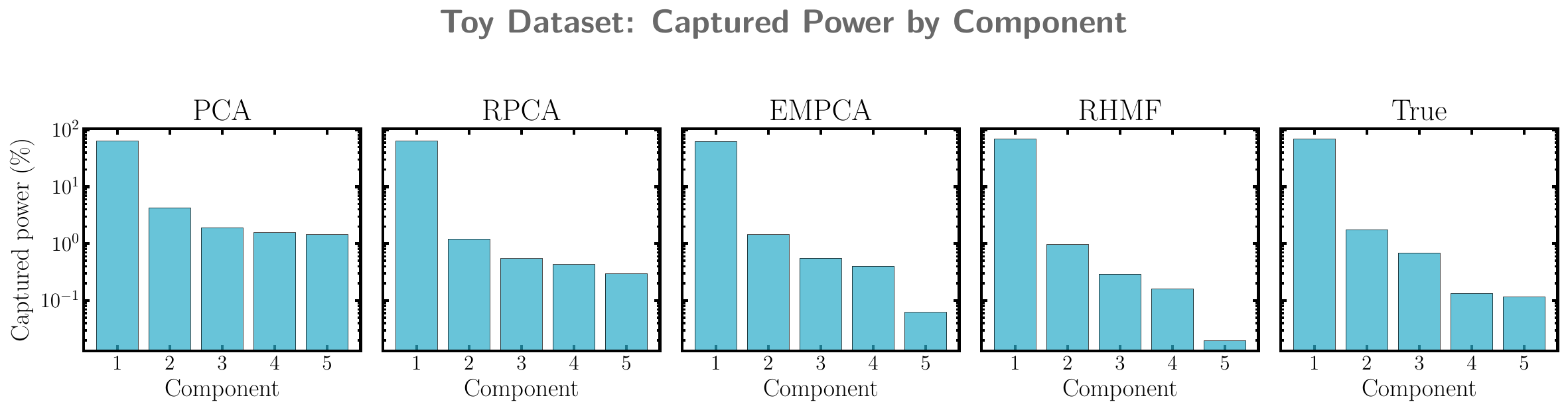}
    \caption{Data power captured by each component for PCA, RPCA, EMPCA, RHMF, and the truth on the toy dataset, as a percentage of the total mean-square data power (note the logarithmic scale). All methods are fit on the training set and evaluated on the full dataset in the canonical frame of Figure~\ref{fig:eigenspectra}; the true panel shows each generative component's contribution $\mathbb{E}[a^2]\,\|g\|^2$. PCA inflates components 2--5 well above the true values because they absorb noise and outlier power; RPCA and EMPCA are closer; RHMF most closely tracks the true fall-off.}
    \label{fig:explained_variance}
\end{figure}

\begin{figure}
    \centering
    \includegraphics[width=\linewidth]{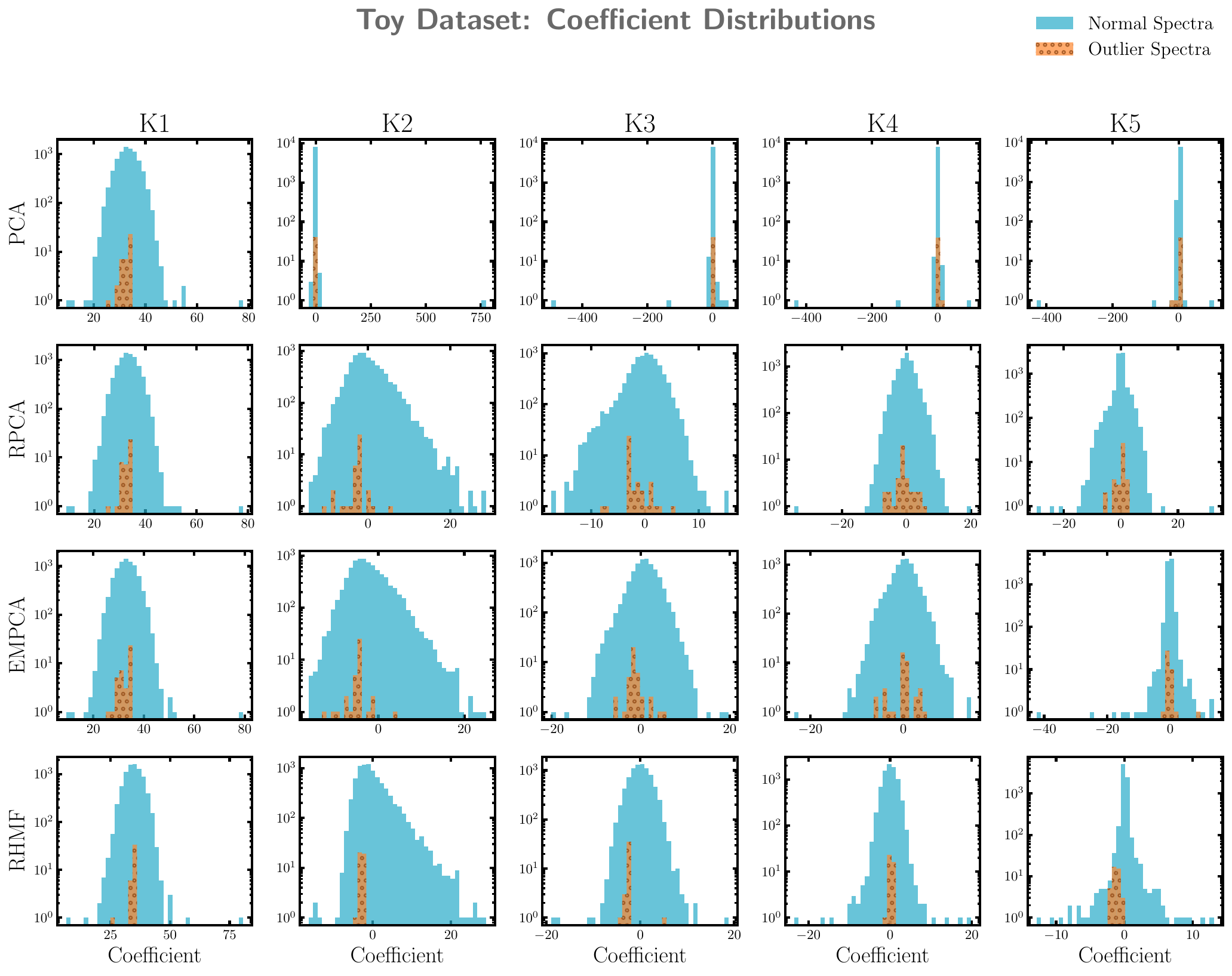}
    \caption{Per-component coefficient distributions for PCA, RPCA, EMPCA, and RHMF on the toy dataset, split into normal (blue) and true outlier (orange, hatched) spectra, with logarithmic counts. For PCA, a handful of corrupted spectra acquire extreme projection coefficients that stretch the axes by an order of magnitude, illustrating how outliers dominate its coefficient space. For the other methods the coefficient distributions are well-behaved, but the outlier spectra sit \emph{within} the bulk of the normal population in every component: coefficient values alone do not separate these outliers. This motivates the robust weights (Figure~\ref{fig:toy_weights}) as the primary outlier diagnostic, with coefficients playing a complementary role (see Discussion).}
    \label{fig:coefficients}
\end{figure}

To understand how RHMF compares with other matrix factorization methods, Figure~\ref{fig:eigenspectra} shows the learned basis vectors for PCA, RPCA, EMPCA \citep[weighted expectation-maximization PCA;][]{bailey2012}, and RHMF side-by-side with the ground truth basis from the synthetic data generation. EMPCA is the appropriate weighted baseline: like RHMF it uses the inverse-variance weights and handles missing data, but it has no outlier mechanism. Figure~\ref{fig:explained_variance} compares the data power captured by each component, showing that PCA inflates its trailing components by absorbing noise and outlier power, while RHMF most closely tracks the true fall-off. Figure~\ref{fig:coefficients} shows the per-component coefficient distributions split into normal and outlier spectra, addressing whether outliers can be identified from coefficient values alone. Appendix~\ref{sec:toy_model_details} additionally shows how the recovered RHMF eigenspectra vary across the $(K, Q)$ hyperparameter grid (Figures~\ref{fig:eigenspectra_vs_q} and \ref{fig:eigenspectra_vs_k}).

We performed cross-validation for the fitted RHMF model using the KL-divergence specified in Section~\ref{sec:cv}.
The validation score for each hyperparameter pair $(Q, K)$ is shown in the top panel of Figure~\ref{fig:cv}.
The lowest divergence is achieved at $Q=5$, and at the true rank of $K=5$; this is the pair used in Figures~\ref{fig:toy_residuals} and \ref{fig:toy_spectra}. The score is more sensitive to the choice of $Q$ than $K$, as the $K>5$ cases with $Q=5$ fare similarly.
$K<5$ performs poorly, reflecting the inability of the model to capture all the common structure in the data.
The optimal $Q=5$ implies that the transition from the quadratic to logarithmic penalty in the residuals occurs near $5\sigma$ from zero. 
At all fixed $K$, there is a local optimum in $Q$, and that optimum shifts to larger $Q$ as $K$ increases.
This likely reflects the fact that the $K<5$ models, which are not flexible enough to capture all the common structure, can compensate by downweighting more aggressively (smaller $Q$) such that the low-rank structure of interest is no longer captured in the inferred basis.
Luckily, it is relatively easy to diagnose this issue by looking for shared structure in the residuals across many spectra.

In the bottom two panels of Figure~\ref{fig:cv} we assess the ability of the model to automatically identify outliers on both a per-pixel and a per-object basis, using the inferred robust weights.
To do this, we set a hard threshold for each the pixel-level robust weights $w_{ij}^{\rm robust}$ and the object-level weights $w_i^{\rm object}$.
Individual pixels are considered outliers if $w_{ij}^{\rm robust} < 0.5$, and whole spectra are considered outliers if $w_i^{\rm object} < 0.9$, where $w_i^{\rm object}$ is the median along $j$ as defined in Eq.~\eqref{eq:w_spec}.
Then, we calculate the F1 score for each case, where the F1 score is a common metric for binary classification tasks that combines both precision and recall.
These quantities are defined as
\begin{align}
    \mathrm{precision} &= \frac{\mathrm{true~positives}}{\mathrm{true~positives} + \mathrm{false~positives}}, \\
    \mathrm{recall} &= \frac{\mathrm{true~positives}}{\mathrm{true~positives} + \mathrm{false~negatives}}, \\
    \mathrm{F1} &= \frac{2 \cdot \mathrm{precision} \cdot \mathrm{recall}}{\mathrm{precision} + \mathrm{recall}}.
\end{align}
An F1 score of 1 indicates perfect classification; no false positives nor false negatives.

Both the F1 panels reveal similar behaviour: the outlier identification is relatively \emph{insensitive} to $Q$ and $K$, provided that both are sufficiently large.
The models with $K<5$ struggle in that they cannot capture the underlying structure with a true rank of $5$, leading to false positives.
This can be partially remediated by increasing $Q$ to downweight less aggressively, but it is clearly not ideal.
$Q<3$ causes false positives despite large enough $K$, because the downweighting is too aggressive.
These results demonstrate that the best outlier identification is obtained near the best validation score, but also that picking $K$ or $Q$ too large does not dramatically degrade the outlier identification.
However, this is likely contingent on the accuracy of the provided measurement uncertainties, as well as the degree to which the outliers do not share low-rank structure with the rest of the data.

The differences between these three quality metrics reflect their distinct objectives and modeling assumptions. The KL divergence (top panel) measures generalization performance on held-out data: how well the model compresses the full data distribution into a sufficient low-rank structure. In contrast, the F1 scores measure the accuracy of automated outlier detection against \emph{known ground-truth labels}. These objectives are related but not identical: a model can generalize well without achieving optimal outlier detection at every resolution, and vice versa. The KL-optimal hyperparameters ($K \geq 5$, $Q=5$) prioritize capturing the underlying common structure of normal spectra with conservative downweighting, which typically balances both objectives. However, aggressive downweighting (smaller $Q$) can achieve better per-pixel F1 by catching isolated pixel-level outliers more effectively, while the per-object F1 metric suggests intermediate $(K, Q)$ values that trade off sensitivity for specificity differently. Each metric weights different aspects of the model's behavior: generalization on inliers (KL), detection of localized anomalies (per-pixel F1), and classification of entire spectra (per-object F1).

The good news is that the practical impact of this variation is modest. The landscape across the parameter space where $K \geq 5$ and $Q \gtrsim 3$ is relatively flat for all three metrics, indicating genuine robustness to reasonable hyperparameter choices within this regime. For practitioners, we recommend using the KL-based validation score, as it provides a principled model-selection criterion that does not require labeled ground truth and naturally avoids over-fitting when $K$ is not too large and $Q$ is not too aggressive. The toy example confirms this choice is sensible: the KL-optimal hyperparameters yield outlier identification performance close to the F1 optima. Moreover, the insensitivity of the method to over-fitting when both $K$ and $Q$ are sufficiently large suggests that cross-validation alone is often a reliable guide. For users with access to labeled outlier data in their own domain, the F1 panels can be computed post-hoc to verify that the chosen hyperparameters suit their specific outlier morphologies; however, this is a diagnostic check rather than a prerequisite or necessity.

\subsection{Outliers in the Gaia RVS main sequence} \label{sec:gaia}

To demonstrate RHMF in a more realistic setting, we applied it to Gaia/RVS spectra to look for unusual main-sequence stars \citep{collaboration2016,vallenari2023,katz2023}. Here, \emph{unusual} is qualitatively defined by the spectrum deviating significantly from typical stars in neighboring color-magnitude space, and quantified by the inferred robust weights. Examples classes could be binary systems, stars with peculiarly high rotation, stars with signatures of accretion or chromospheric activity, or perhaps other unknown unknowns.

The Gaia RVS spectra cover a wavelength range of 846--870\,nm in vacuum wavelengths, sampled at 2401 pixel locations corresponding to a sampling of approximately 0.01\,nm per pixel. The spectra are described in \citet{katz2023}, and we preprocessed them by clipping 40 pixels from the edges of each spectrum to remove regions with reduced flux quality, leaving 2321 usable pixels per spectrum. For each spectrum, we used reported flux uncertainties and computed the inverse-variance weights $w_{ij}^{\rm data} = 1/\sigma_{ij}^2$, which encode both the measurement uncertainty and spectral quality information in our analysis.

\begin{figure}
    \centering
    \includegraphics[width=0.88\linewidth]{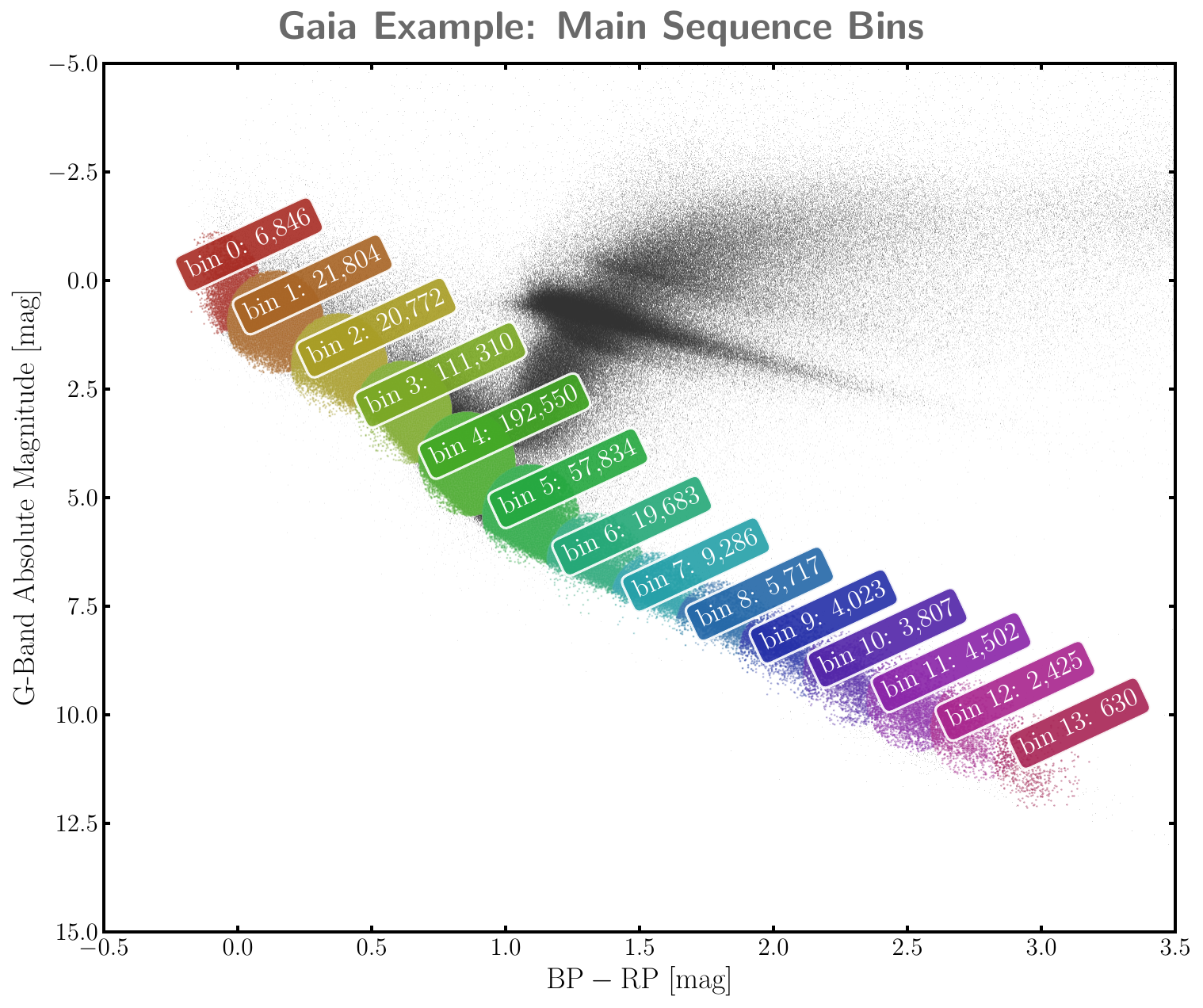}
    \caption{Gaia BP$-$RP color vs absolute G magnitude diagram, showing all stars in black. The subset we used for the analysis are colored by the bin they belong to. Each bin is labelled by an index 0 through 13, and the number of spectra in each bin is also shown. The bin edges overlap, and so many spectra belong to two bins, although this is not shown in the plot.}
    \label{fig:hr_bins}
\end{figure}

We ran RHMF on stars in local regions of color-magnitude along the main sequence (Figure~\ref{fig:hr_bins}).
The motivation for local regions is largely interpretability, in that we want to be able to understand the common structure in the spectra of stars in a particular region of the Hertzsprung-Russell diagram, and then identify spectra that deviate from that common structure.\footnote{An added benefit to running RHMF on local regions is computational efficiency.}
We used 14 regions, evenly spaced in Gaia BP$-$RP color and nearly evenly spaced in G-band absolute magnitude.
The widths in color and magnitude were chosen to ensure that they slightly overlap, such that some stars belong to two bins. 
Within each bin, we randomly selected half the spectra to fit using RHMF, where we initialised the vectors with SVD and iterated until the maximum fractional g-step adjustment was $<10^{-4}$.
We chose to use $K=10$ basis vectors and set $Q=5$ for all bins. We did not perform any cross-validation, primarily because we are interested in identifying \emph{potentially} interesting spectra (which will be vetted by humans), and so the exact classification performance is not of primary importance to us.
And while we recommend cross-validation in practice, it is worth noting that our toy example showed that outlier identification is relatively insensitive to the choice of $Q$ and $K$, provided that they are not too small.
The vectors fit from the randomly selected half of the spectra in each bin were then used to predict coefficients and weights for the other half.

\begin{figure}
    \centering
    \includegraphics[width=0.76\linewidth]{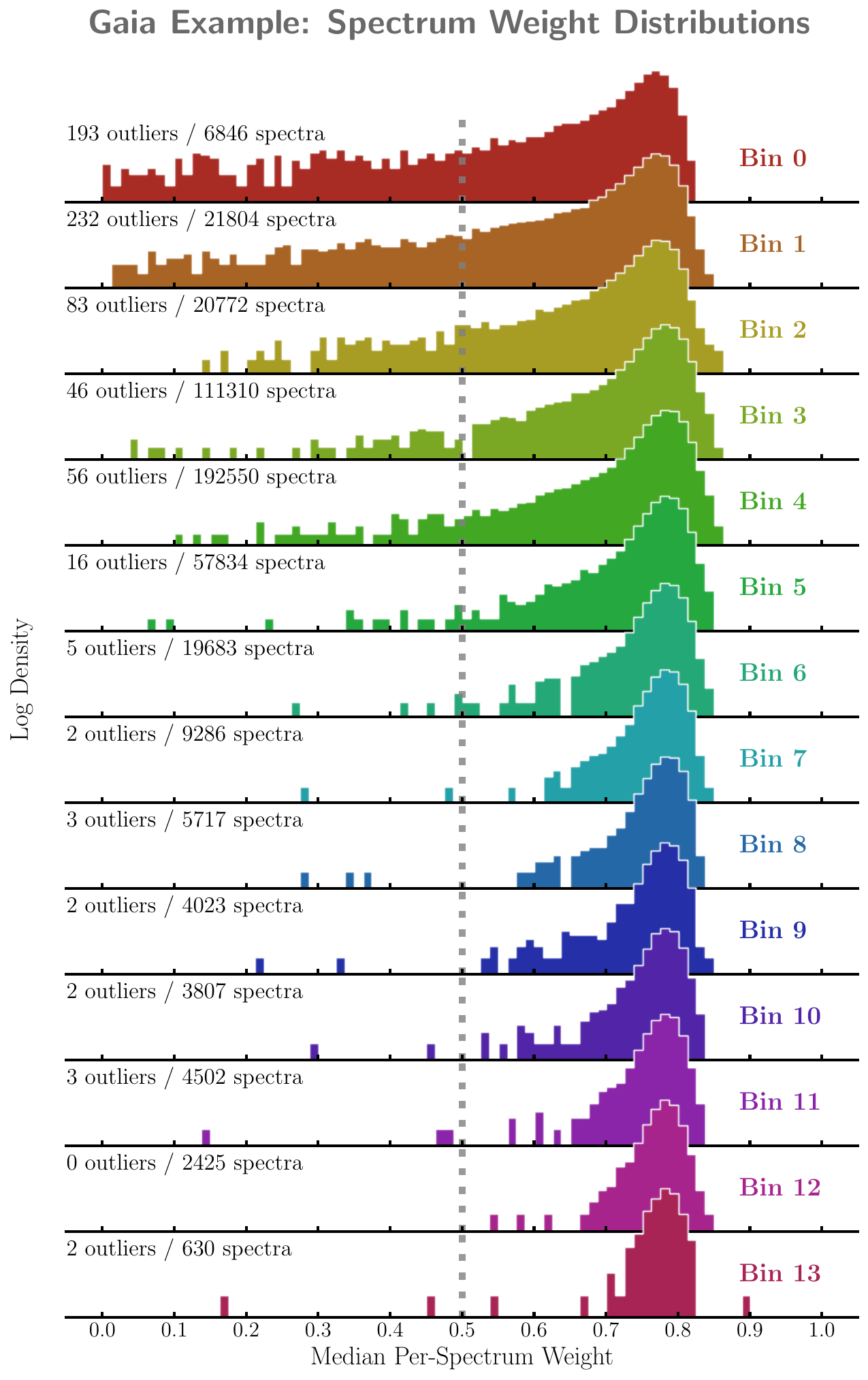}
    \caption{Distribution of the object-level weights $w_i^{\rm object}$ for each bin, with a vertical grey dashed line showing the threshold of $0.5$ for outlier identification. The number of identified outliers with $w_i^{\rm object} < 0.5$, as well as the total number of spectra in each bin, is labelled next to the corresponding histogram.}
    \label{fig:gaia_weights}
\end{figure}

With basis coefficients, predicted model spectra, and robust weights for all spectra in each bin, we were then able to identify outlier spectra as those with low per-object weights.
Here, the per-object weights $w_i^{\rm object}$ were calculated as the lowest 1\% percentile over $j$ from the per-pixel robust weights $w^\mathrm{robust}_{ij}$.
We set a threshold of $0.5$, such that if $w_i^{\rm object} < 0.5$ then spectrum $i$ is flagged as a potential outlier, this is shown as a vertical dashed line in the figure.
The number of identified outliers is labelled for each bin, as well as the total number of spectra in that bin.

For the lower main-sequence bins (K and M-type stars; bins 6 through 13) we see similar behaviour in the distributions of weights as in the toy example, with a large population of spectra with large weights (inliers), and a very sparse tail of spectra with weights ranging from $\lesssim 0.5$ to $\lesssim 0.1$ (outliers).
In the upper main sequence (closer to A and F-type stars; bins 0 through 5), the tail of the distribution is better populated, making it more difficult to identify a well-defined outlier separation.
This is not completely explained by number counts: there are 193 outliers marked in bin 0 from 6,846 spectra; and just 2 outliers in bin 7 from 9,286 spectra.
It is consistent with the fact that the upper main sequence is more spectrally diverse, and so there are more spectra that deviate from the common structure in the data, and thus more spectra with low weights.
We note, however, that this is not the only interpretation: the variation in flagged-outlier rate is also consistent with the threshold $w_i^{\rm object}<0.5$ not corresponding to a calibrated false-positive rate, with the true rate varying across bins simply because the distribution of $w_i^{\rm object}$ varies with the underlying spectral diversity. Disentangling ``the upper main sequence has more genuinely peculiar stars'' from ``our threshold is differently strict in different bins'' requires the injection-recovery experiment mentioned above; we cannot resolve it from the current data.
Naturally, having more components (increasing $K$) could capture common kinds of anomalies (e.g., emission lines).
However, this introduces a double-edged sword: a flexible model can capture more structure in the data, but some of that structure may be capturing relatively common astrophysics! 
This emphasizes a fundamental limitation: if the objects you consider outliers share some low-rank structure between them, then higher $K$ components will capture that structure. Ultimately the investigator must decide on the balance between flexibility and interpretability.

While the toy example demonstrated that outlier identification is relatively insensitive to the choice of $Q$ and $K$ provided both are sufficiently large, real stellar spectra exhibit several important differences. The heterogeneity of spectral types in the real data is considerably greater: the upper main sequence (hotter A and F-type stars) exhibits substantially more spectral diversity than the cooler lower main sequence. This manifests as varying flagged-outlier rates across the Hertzsprung-Russell diagram (e.g., 193 outliers from 6,846 spectra in the hottest bin versus only 2 from 9,286 spectra in a cooler bin), reflecting the genuine diversity of stellar spectra. However, the ranking of models by cross-validation score remains stable across modest changes in both $K$ and $Q$ (see Section~\ref{sec:cv}), suggesting that using fixed hyperparameters across all bins is reasonable for identifying candidate outliers for subsequent human vetting.

\begin{figure}
    \centering
    \includegraphics[width=0.98\linewidth]{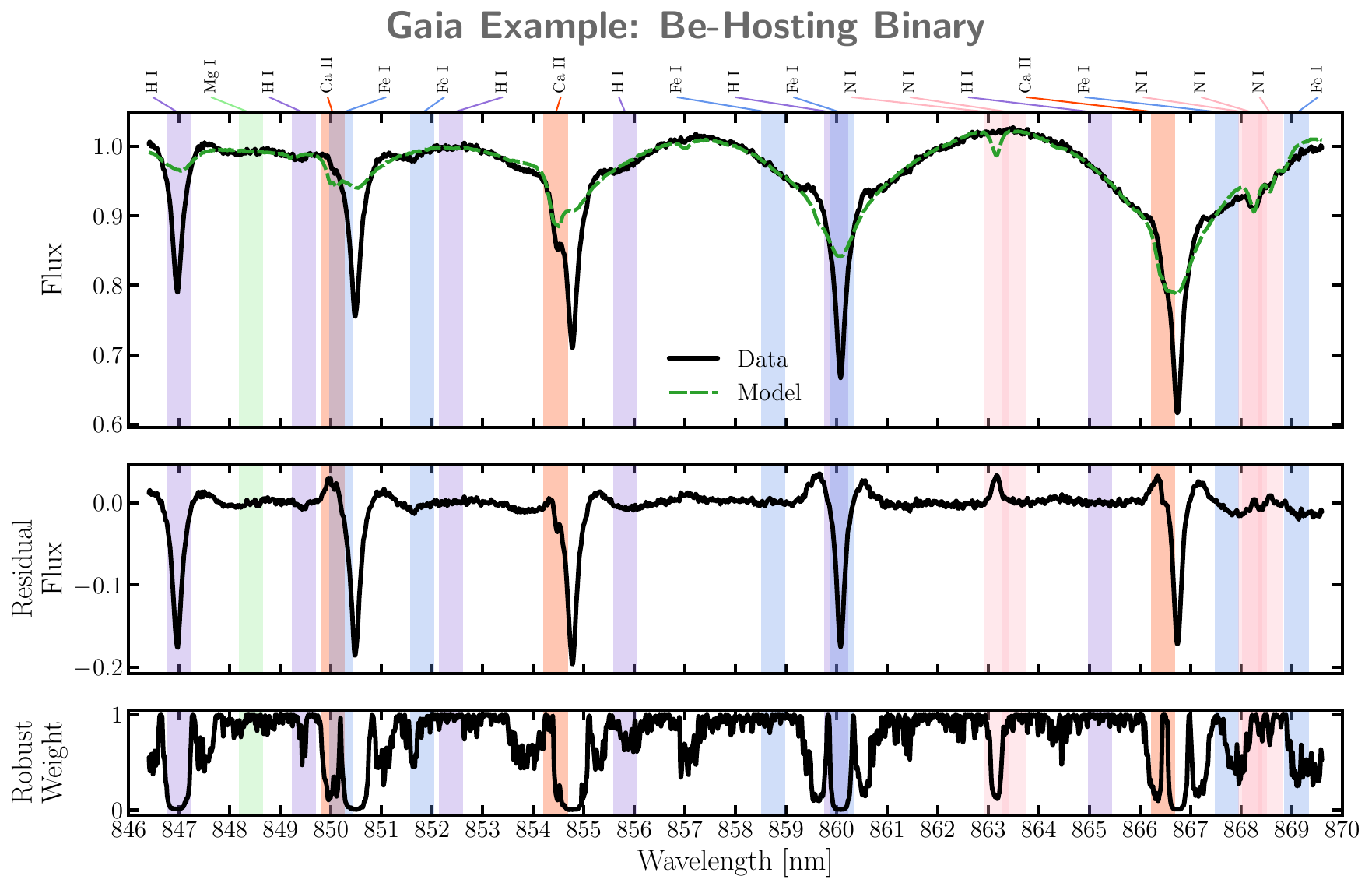}
    \caption{Gaia/RVS spectrum (black) of Gaia DR3 1363284299777747584 (HD 162732; 88 Her) that we identified as an outlier ($w_i^{\rm object} = 9\times10^{-3}$) in bin 0, shown with the RHMF reconstruction (green, dashed). This system is a suspected binary system that hosts a Be star with an equatorial disk of gas.}
    \label{fig:gaia_spec_1}
\end{figure}

\begin{figure}
    \centering
    \includegraphics[width=0.98\linewidth]{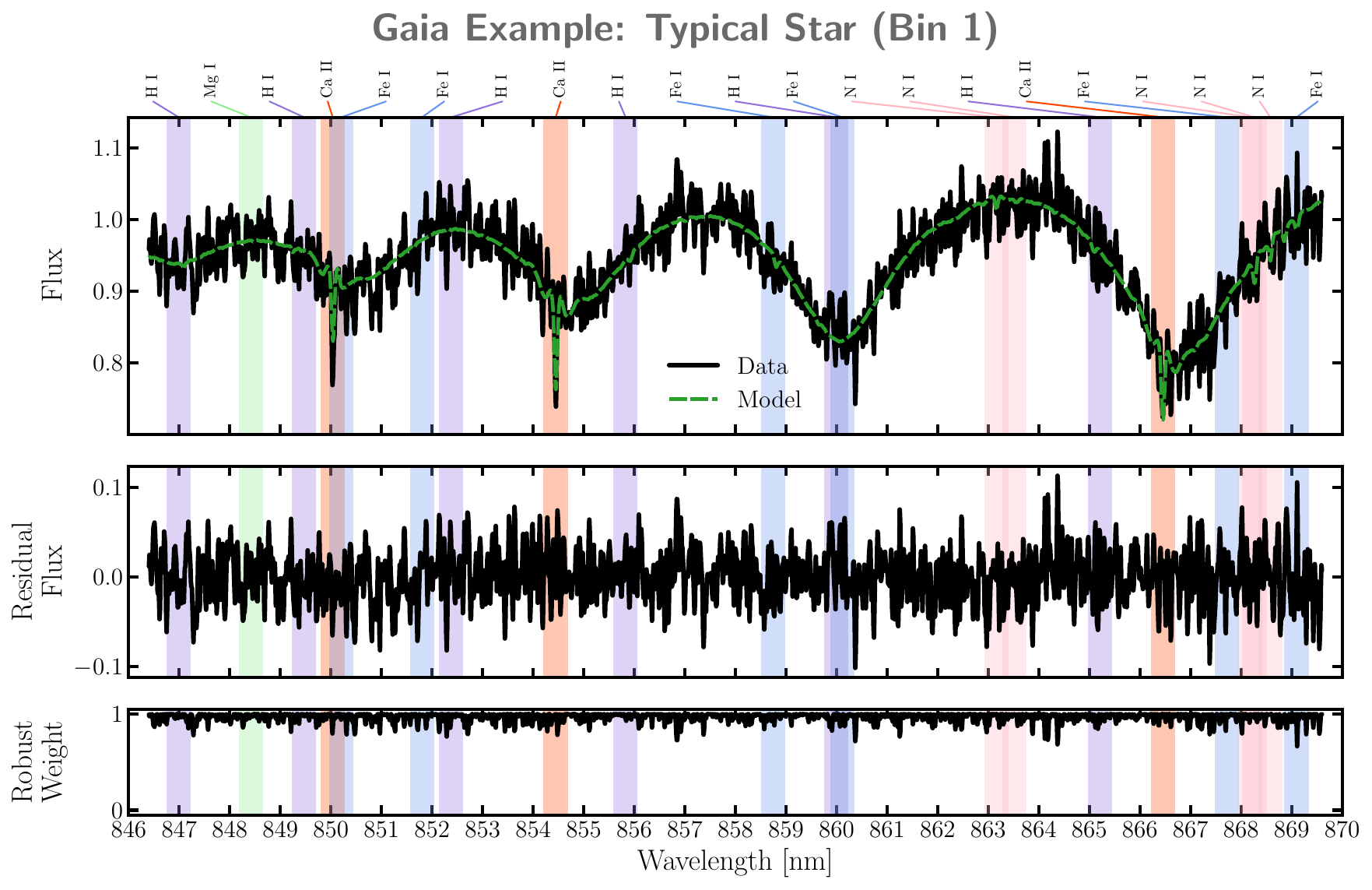}
    \caption{Gaia/RVS spectrum (black) of a more typical star (Gaia DR3 5309096898078973568) on the upper main sequence, shown with the good reconstruction (green, dashed). Unlike the outlier spectrum shown in Figure~\ref{fig:gaia_spec_1}, this spectrum has a high object-level weight of $w_i^{\rm object} = 0.985$, consistent with it being a typical star in this bin.}
    \label{fig:gaia_spec_normal}
\end{figure}

As a method-confidence check, we first verified that RHMF assigns very low object-level weight to a spectrum that is obviously off the main-sequence manifold. 
The lowest-weight object in bin 0, shown in Figure~\ref{fig:gaia_spec_1}, is 88 Her (HD 162732; Gaia DR3 1363284299777747584) with $w_i^{\rm object} = 9\times10^{-3}$. The reconstruction is poor, as expected for a spectrum with broad Paschen lines indicative of a hot, fast-rotating star (cf. Figure~\ref{fig:gaia_spec_normal}). 
This system is a known suspected binary hosting a Be-shell star with an equatorial disk of ejected gas viewed edge-on \citep{abt1995,rivinius2006}; the classification is taken from the literature and is not a finding of this work. 
We include it solely to demonstrate that the method does the obvious thing on an obvious case.

\begin{figure}
    \centering
    \includegraphics[width=0.97\linewidth]{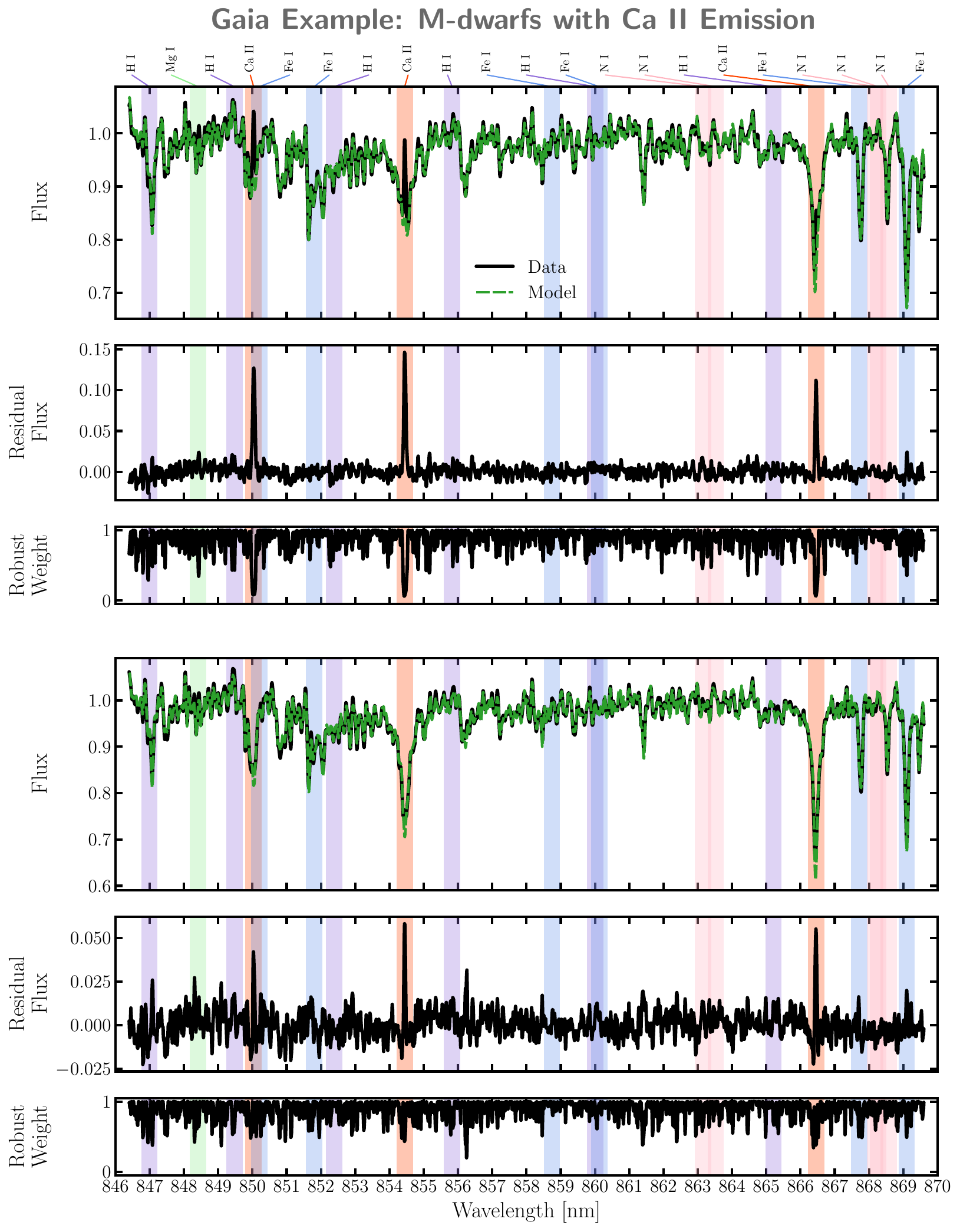}
    \caption{Two M-dwarf outliers identified in bin 13: Gaia DR3 3136952686035250688 (top), and Gaia DR3 3195919254111314816 (bottom). Both show the model reconstructions in green (dashed), where the residual panels highlight emission in the Ca II triplet, consistent with stellar activity. The emission in the top panel is visibly obvious. The emission signature in lower panel is only discernible when compared to the model reconstruction of normal spectra, highlighting the utility of RHMF for identifying subtle outliers among large heteroskedastic datasets.}
    \label{fig:gaia_spec_2}
\end{figure}

A more compelling demonstration is in bin 13 (the M-dwarf bin), shown in Figure~\ref{fig:gaia_spec_2}. Both flagged objects show emission in the Ca II triplet, consistent with chromospheric activity or accretion. In the top panel (Gaia DR3 3136952686035250688; $w_i^{\rm object} = 0.173$) the emission is conspicuous to the eye and would likely be caught by a sufficiently careful by-hand inspection of the bin. The lower panel (Gaia DR3 3195919254111314816; $w_i^{\rm object} = 0.453$) is less apparent: the emission is sub-percent in depth, is comparable to the level of per-pixel uncertainties, and is essentially invisible in the raw spectrum. Yet it stands out cleanly once compared against the RHMF reconstruction built from the surrounding M-dwarf population. This is precisely the regime in which RHMF should be most useful: subtle, line-localised deviations that a fixed-template or PCA-based search would either average over or absorb into the basis. Both examples here are presented as case studies, not as a population-level result; a systematic characterisation of anomalous populations identified by RHMF is a natural extension of this work.

\subsection{Scaling to the full Gaia RVS sample} \label{sec:gaia_full}

The experiment above demonstrated that RHMF can be an effective model even when restricting to local subsets of the data (e.g., in color-magnitude bins) and when skipping cross-validation and using sensible hyperparameters.
Here we do the opposite on both counts: we fit a single RHMF model to \emph{every} Gaia RVS spectrum, and we choose $(K,Q)$ by the held-out score of Section~\ref{sec:cv}. In doing so we will demonstrate that the algorithm scales to a full survey without partitioning the data, and that it exposes a subtlety in relying on the KL-divergence as the only cross-validation metric, which was not apparent in the toy model. 

Here our sample includes 999,645 stars in Gaia DR3 that have an RVS spectrum.
Unlike the binned experiment, no color or magnitude selection is applied: the upper main sequence, the giant branch, the red clump and the white dwarf sequence are all fit by one basis. We use the same algorithm presented earlier, but with a few computational modifications to accommodate the larger data volume. Specifically, we evaluate the a- and g-steps in row-blocks, accumulating the $K\times K$ and $K\times M$ normal-equation sums block by 
block, and allow for data sharding across multiple GPUs. We use the same stopping criterion as before: requiring a maximum fractional g-step adjustment of $<10^{-4}$.

\begin{figure}
    \centering
    \includegraphics[width=0.5\linewidth]{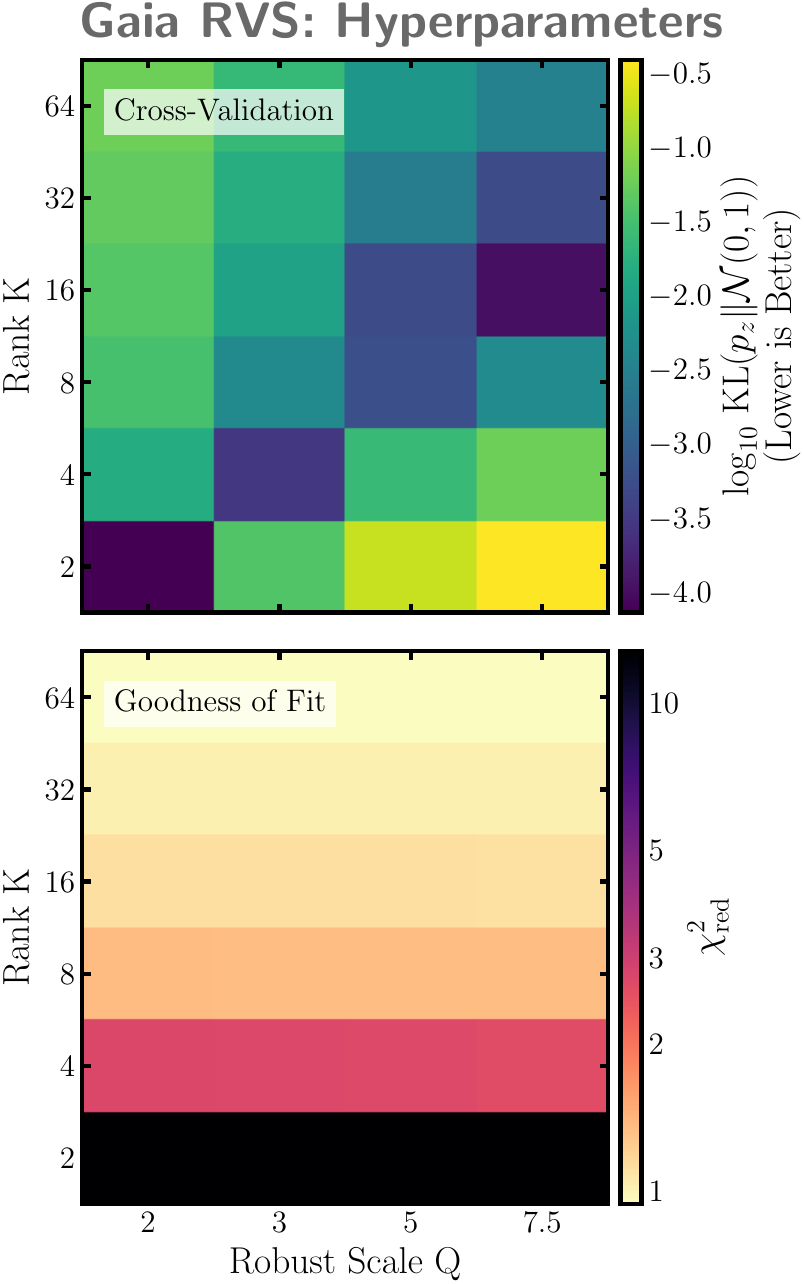}
    \caption{Hyperparameter selection for the full Gaia RVS sample. \emph{Top:} the cross-validation KL divergence from the distribution of standardised residuals $z_{ij}$ on held-out data to $\mathcal{N}(0,1)$, over a grid of rank $K$ and robust scale $Q$. \emph{Bottom:} the reduced chi-squared of the same models on the same held-out data. Note the logarithmic colorbar for $\chi_\textrm{red}^2$ used for clarity. Strictly the KL divergence is lowest for $K=2$ and $Q=2$, but this is an inflexible model ($\chi_\textrm{red}^2 \approx 12$). The adopted hyperparameters are $K = 16$ and $Q = 7.5$: the second-lowest KL divergence measured, with $\chi_\textrm{red}^2 \approx 1$.}
    \label{fig:full_rvs_cv}
\end{figure}

We cross-validated over $K \in \{2,4,8,16,32,64\}$ and $Q \in \{2,3,5,7.5\}$ on 10\% of the data: half of that sample (5\% of the full sample) was used for fitting, and the other half for scoring. The resulting scores are shown in the top panel of Figure~\ref{fig:full_rvs_cv}. The minimum sits in a diagonal valley: as $K$ increases, the $Q$ that minimises the score increases with it, as expected, since a more flexible model leaves smaller residuals and requires less aggressive downweighting to make $z_{ij}$ resemble a unit normal.

The global minimum of the KL-divergence is found at $K=2$ and $Q=2$, but this is not a sensible model for us to adopt. The reduced chi-squared on the held-out data is $12.7$ (bottom panel of Figure~\ref{fig:full_rvs_cv}): highlighting that a rank-2 basis is far too inflexible to capture the diversity of the full RVS sample. The score is nonetheless near-zero because the two quantities it depends on, $\hat{\mu}_z$ and $\hat{\sigma}_z$, are computed \emph{after} robust downweighting. When the model is badly rank-deficient and $Q$ is small, almost every pixel acquires a small $w^{\rm robust}_{ij}$, the surviving $z_{ij}$ are driven toward zero variance, and $\hat{\sigma}_z$ can be tuned through unity on the way down. The score is then satisfied not because the model explains the data, but because it has declared most of the data to be outlying.

While we still recommend the KL-divergence (Equation~\ref{eq:score}) as the preferred measure during cross-validation, this situation is worth highlighting because it illustrates that the divergence should not be taken blindly or in isolation. KL-divergence measures calibration, not model adequacy. It asks whether the residuals that remain look like the noise model says they should, and a model that discards enough data can always answer yes. In the toy example of Section~\ref{sec:toy} the issue does not arise, because the true rank is small and even the smallest $K$ on the grid is close to adequate; it emerges here precisely because the full RVS sample is far more heterogeneous than any single bin. It is recommended that the KL-divergence be interpreted alongside some goodness-of-fit statistic, or regarded as a criterion for choosing $Q$ at a fixed $K$, with a separate decision-making process used to determine a sufficient $K$.

Applying that rule to Figure~\ref{fig:full_rvs_cv} leaves $K=16$, $Q=7.5$, which has the second-lowest score on the grid ($1.0\times10^{-4}$, against $7.2\times10^{-5}$ for the degenerate rank-2 model) and $\chi^2_{\rm red} = 1.10$. Higher ranks fit better still in the chi-squared sense, but $K=64$ reaches $\chi^2_{\rm red} = 0.94$, below unity, indicating that the basis has begun to absorb noise; and, as noted in Section~\ref{sec:gaia}, a more flexible basis will increasingly absorb the very anomalies we are trying to isolate. We adopt $K=16$, $Q=7.5$ and refit on all $999{,}645$ spectra. 
Convergence was reached after 70 iterations, which took nine minutes on an NVIDIA RTX A6000 using 64 bit floating-point arithmetic. The in-sample diagnostics of that final fit ($\hat{\sigma}_z = 1.009$, $\chi^2_{\rm red} = 1.11$) are consistent with the held-out values, as expected when $N \gg K$.

\begin{figure}
    \centering
    \includegraphics[width=\linewidth]{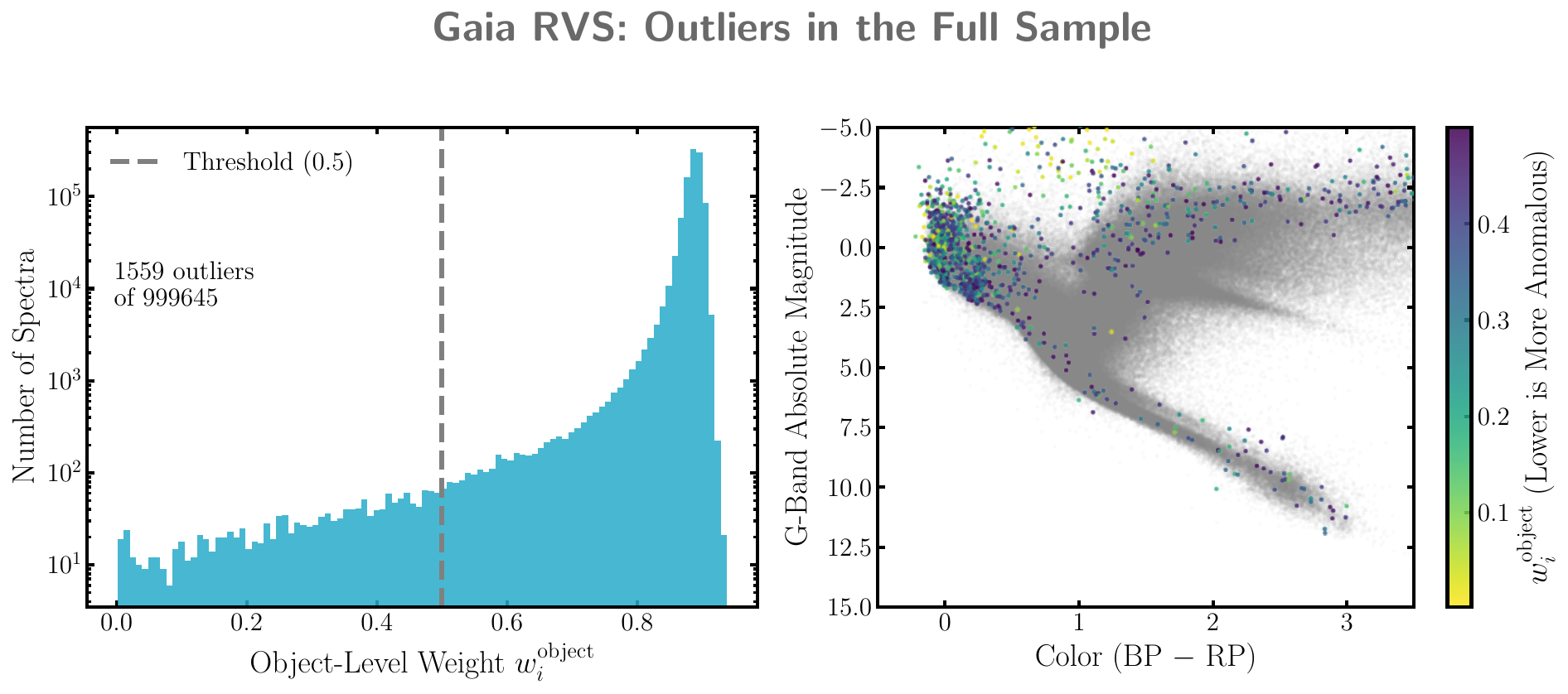}
    \caption{Outliers in the full Gaia RVS sample at $K=16$, $Q=7.5$. \emph{Left:} the distribution of the object-level weight $w_i^{\rm object}$, here the 1\% quantile over pixels of the per-pixel robust weights, with the $0.5$ threshold marked. \emph{Right:} the color-magnitude diagram, with the full sample in grey and the 1{,}559 flagged spectra colored by $w_i^{\rm object}$.}
    \label{fig:full_rvs_outliers} 
\end{figure}

Using the same object-level statistic and threshold as before ($w_i^{\rm object} < 0.5$, with $w_i^{\rm object}$ the 1\% quantile over pixels), the model flags $1{,}559$ spectra, or $0.16\%$ of the sample. Their distribution is shown in Figure~\ref{fig:full_rvs_outliers}. The bulk of the population sits near $w_i^{\rm object} \approx 0.9$ with a long tail toward zero, and the flagged spectra are concentrated toward the hot, luminous end of the color-magnitude diagram: of the 77 spectra scoring below $0.05$, 69 lie blueward of BP$-$RP $=1.5$ and 72 are brighter than $M_G = 0$.

We visually inspected all $1{,}559$ spectra flagged as outliers and grouped them by similarity in their spectrum residuals. While we avoid interpreting every flagged outlier (or groups of outliers), it is worth commenting on the most frequent outlier types identified. The most anomalous group (lowest object score) are blue supergiants, and the largest group includes hot stars with broad line profiles, where line asymmetry or emission in the Ca II triplet is the dominant residual feature. Other dominant groups include: cool stars with chromospheric activity; stars with significant interstellar absorption; Mira variables; Long Period Variables; or stars rich in neutron-capture abundances. A gallery of these examples is shown in the Appendix (Figures~\ref{fig:tax_bsg} to \ref{fig:tax_ncap}), and in the supplementary materials we provide a catalog of all $999{,}645$ Gaia DR3 spectra, their associated metadata, best-fitting model spectra, and robust weights. This catalog can be used to readily identify outliers in the Gaia RVS sample, and to construct matched-filter searches for specific classes of outliers.

\section{Discussion} \label{sec:discussion}

We have introduced a generalization of PCA, which we call robust heteroskedastic matrix factorization.
RHMF has all the nice properties of both HMF \citep{tsalmantza2012} and robust regression \citep{holland1977, hoaglin1983}, in that it can handle missing data, heteroskedastic noise, and outliers, all in a principled way.
The usefulness of a model of this kind is two-fold: one may care about the low-rank structure without being burdened by outliers, or one can use the inference to identify outliers and then investigate them further.
We demonstrated the former in the toy example, and the latter in the Gaia/RVS example.
Like all models, it has strengths (hopefully demonstrated in the preceding sections), and shortcomings, which we will discuss below.


In PCA the number of basis vectors $K$ can be determined after fitting a model by looking at the variance explained by each component, and picking a cutoff threshold. In RHMF, this is not possible: the number of basis vectors $K$ must be determined before fitting the model. The user must also decide on the soft outlier threshold $Q$, which controls the aggressiveness of down-weighting outliers.
We advocated for a data-driven, cross-validation approach to selecting these hyperparameters through assessing their predictive performance on held-out data.
Typically, this might be done through some kind of reconstruction loss: a summary statistic over the uncertainty-weighted residuals between the model prediction on held-out data. 
In general more basis vectors $K$ will always do better because the model is more flexible, so one must decide when they have \emph{enough} basis vectors, which is not a well-defined criterion.
But there are additional nuances to cross-validation in this context. 
For example, when evaluating performance on held-out data, should we use only the inverse variance of the data, or should we include the robust weights as well? If we use only the data weights, then increasing $K$ will always do better, because we are accidentally assessing the model's ability to fit the outliers, not ignore them. If we use the total weights (including the robust downweighting), then the model could be downweighting real features, which should be contributing to the performance metric.
For these reasons, we advocated a practical approach where we measure how closely the weighted residuals on the held-out data follow a unit normal distribution (Eq.~\eqref{eq:score}). The key insight is that, under a well-calibrated model, applying the data precision weights and the inferred robust weights together should make the residuals look like draws from $\mathcal{N}(0,1)$: the data weights account for known measurement noise, while the robust weights shrink outlier residuals toward zero. We therefore score models by the KL divergence from the empirical distribution of these weighted residuals to $\mathcal{N}(0,1)$, with smaller values indicating a better fit to the held-out data.

In the toy model, we assessed the outlier identification performance with the F1 score over a grid of hyperparameter values.
Comparing with the cross-validation results, the F1 score per-pixel is maximized by the hyperparameters with the best validation score, and the per-object maximum is nearby.
This is reassuring as it indicates that the validation metric is a sensible choice.
Interestingly, the F1 scores only decrease slightly at larger $Q$ where the validation score degrades.
It is plausible this is because outliers truly are sparse and without shared low-rank structure, leading to good performance without aggressive down-weighting, but it is unclear if this generalises to realistic datasets where outliers may not satisfy those assumptions. 
For example, consider the case where in addition to the main objects of interest making up the majority of the dataset with shared low-rank structure, there exists a second class of rare objects that constitute a small percentage of the population, which also share low-rank structure.
Lower $K$ will prioritise common objects and identify the second class as outliers, but larger $K$ will additionally fit the second class instead of downweighting them.
Whether or not the second class are ``outliers'' is in the eyes of the beholder, but setting $K$ adequately for either view will require careful judgement. 
In any case, the model still has utility, it simply requires care in interpretation.


The degree to which the robust weights can be used to identify outliers depends on the investigators' beliefs about the data, and the model assumptions.
In situations where the uncertainties are well-known and truly Gaussian, and where $Q$ and $K$ have been set carefully by some quantitative method such as cross-validation, it is reasonable to use the population of weights as a means to ascribe a quantitative measure of outlier degree. 
However, these conditions are rarely met in practice. 
Spectra with under-estimated uncertainties can easily manifest as outliers through the robust weights, since they might receive near-universal down-weighting. 
This is somewhat remediated by close inspection of the residuals between the data and the model, as objects that are not genuine outliers (but just have under-estimated uncertainties) should remain well-described on average by the model predictions. 
It is reasonable to use the robust weights as a guide for \emph{where to look} for interesting objects in large heteroskedastic datasets, rather than a \emph{statistical measure} of the degree to which an object is truly an outlier.

As already mentioned, outlier identification will be sensitive to the choice of $K$ if those outliers share some low-rank structure.
By construction this was not a problem in the toy model because we designed the outliers to be random, without shared low-rank structure.
In Gaia/RVS spectra, many stars have slight deviations in their continuum normalisation, or common spectral signatures (e.g., binaries). When these features are shared across many spectra they can be captured by the basis functions, and will not be classified as outliers. 
They can often still be identified after inspecting the model basis functions and noticing that one basis has a particular feature, and then looking at the coefficients for that basis to find the spectra that share that feature. But in the strict classification sense of RHMF, they will not be identified as outliers \emph{solely because} of that low-rank structure.

The robust weights and higher-order basis components represent two complementary mechanisms for identifying anomalies, each sensitive to different failure modes of the low-rank model.
The robust weights $w^{\rm robust}_{ij}$ in Eq.~\eqref{eq:w_robust} operate by directly penalizing large per-pixel residuals, flagging immediate deviations from the model: isolated bad pixels, cosmic-ray hits, or localized instrumental artifacts.
In contrast, higher-rank components (larger $K$) can capture shared anomalous structure---systematic effects that manifest across multiple pixels or objects simultaneously, such as stellar rotation signatures, emission lines present in a subset of stars, or instrumental artifacts affecting specific wavelengths.
The toy example demonstrates this distinction: robust weights downweight the random cosmic-ray-like pixels to near-zero weights (Fig.~\ref{fig:toy_residuals}), while the low-rank structure is preserved by the basis vectors. In realistic spectra, these mechanisms work in tandem: robust weights identify the \emph{where} of anomalies (which pixels are unusual), while higher-$K$ components reveal the \emph{what} (patterns shared across similar objects).

A practical implication is that users should inspect both mechanisms when searching for interesting outliers. Spectra with very low per-object robust weights ($w_i^{\rm object} \ll 0.5$) are flagged by the downweighting mechanism and warrant inspection for isolated defects or unexplained deviations. At the same time, examining the basis vectors and associated coefficients can reveal systematic effects that affect broader populations: for example, in stellar spectroscopy, if basis vector $k$ exhibits broad Paschen-line features, then stars with large coefficients $a_{ik}$ for that basis are likely fast rotators, even if their overall robust weights remain moderate. The M-dwarf emission-line example in Figure~\ref{fig:gaia_spec_2} illustrates this: the lower panel shows subtle emission that would not necessarily produce extremely low weights, yet stands out clearly when compared to the model built from typical stars.

The choice of hyperparameters reflects the user's prior about what constitutes interesting structure. Lower $Q$ (more aggressive robust downweighting) emphasizes isolated anomalies and increases model robustness to outliers, at the cost of potentially missing systematic effects. Higher $K$ increases flexibility to capture shared structure, allowing the model to identify subtle or distributed anomalies via basis vector inspection, but risks absorbing real astrophysical variety into the basis (as discussed in Section~\ref{sec:assumptions}). For heterogeneous datasets, where the population contains multiple stellar types or other coherent sub-populations, users may find that a moderate $K$ (chosen by cross-validation) combined with inspection of residuals and basis vectors provides the most complete picture of anomalies. Indeed, this was the case in Section~\ref{sec:gaia_full}, where a $K=2$ model had the lowest KL-divergence but failed to capture the diversity of the full Gaia RVS sample ($\chi_\textrm{red}^2 \approx 12$), while $K=16$ provided a more balanced representation of the data and revealed both isolated and systematic anomalies. This echoes the guidance in the practical workflow (Section~\ref{sec:discussion}): the decision to accommodate multiple populations within the low-rank model or to treat them as anomalies requires scientific judgment and cannot be resolved by the model alone.

It is worth noting that the Gaia DR3 RVS sample is perhaps one of the worst data sets to use for finding outliers, and it is somewhat of a surprise that there are any outliers at all. Only $\sim$$10^6$ RVS spectra were released in DR3 (of $3.3 \times 10^7$ stars). The RVS sample in DR3 was selected to be a very clean sample of stars, where the decisions for what was included in the catalog are not explicitly described. Many well-studied anomalous stars that are in the RVS magnitude range -- which would have a sufficient number of visits and high enough S/N ratio to be included -- are not in the DR3 RVS sample. For these reasons, we would expect fewer outliers (or unusual spectra) to exist in the Gaia DR3 RVS sample than in a strict magnitude-limited (or colour-limited) sample. While RHMF does still identify anomalous stars in this dataset, the unknown selection function for what made it into the Gaia DR3 RVS catalog makes it impossible to do population studies. In DR4, all Gaia/RVS spectra will be released, making it a more astrophysically interesting test for RHMF.


We assumed Gaussian noise in our toy and realistic applications. 
Non-Gaussian noise that is described by a known and closed form distribution can be handled by changing the likelihood function, but not without complications.
Deriving a weight-updating step (as in Appendix~\ref{sec:proof}) requires a closed form marginal likelihood, as we have in Eq.~\eqref{eq:t_like}.
Changing the likelihood from the Gaussian form we have in Eq.~\eqref{eq:hierachical_like} will destroy the analytic marginalisation that yields a Student-t distribution, for most other choices of likelihood.
This can be remedied by selecting a different prior, such that there still exists a closed-form marginal likelihood.
A sufficient, but not necessary, condition for this is prior conjugacy \citep{andrews1974,diaconis1979,west1987}.
This is exploited in the model presented here, as the inverse gamma distribution is the conjugate prior for a normal likelihood with unknown variances \citep[e.g.][]{bernardo2009bayesian}.


The method also comes along with a scalar objective function, which means that in principle it could be trained with stochastic gradient descent methods and mini-batching, trivially scaling to massive data sets. 
Complications arise when trying to initialise the model because naive SVD requires access to the full dataset, but usually subsampling the data is sufficient. All that matters is that you can learn the basis functions, since the coefficients are completely determined by that. In our case, those basis vectors are orthogonal.  That is not strictly required, but in practice the rotation step (which enforces orthogonality) is necessary during training to suppress degeneracy between ${\bf A}$ and ${\bf G}$. Naturally, orthogonality helps for numerical stability and interpretability, but you are free to do whatever you want. You could give the orthogonality condition to ${\bf A}$ instead, for example.

We could extend the model such that the basis functions do not live on a pixel basis, but rather are continuous functions that are represented with sinusoids, wavelets, or something else.
Using sinusoids with regularisation based on the power spectral density of a kernel would allow the basis functions to be Gaussian Processes. 
As we have shown here, RHMF has utility both in identifying outliers \emph{and} in learning a low-rank representation of heteroskedastic data. Using continuous functions would allow for extremely precise basis functions that could be evaluated at any point in the input space, unrestricted to the pixel grid of the data. This would be particularly useful, for example, for spectral components (e.g., stellar, tellurics) with different relative radial velocities, or for extreme precision radial velocity experiments where \emph{very-}sub-pixel precision is required.


The RHMF algorithm presented here is intended for the common astronomical survey setting in which one has a large collection of nominally similar objects, per-feature uncertainties that are of variable (and potentially imperfectly known) quality, scattered missing entries, and a small fraction of `anomalous' objects or features.
RHMF simultaneously serves two purposes: it recovers a low-rank description of the bulk of the data, which is (ideally) undistorted by bad data or anomalies; and, RHMF provides per-feature and per-object measures of data that depart from that low-rank description.

We recommend the following workflow as a practical default, that may be useful in many applications.
(1) Initialise the basis with a (possibly randomized) singular-value decomposition of the data.
This is computationally inexpensive and is often adequate even when computed from a representative subsample.
(2) In the absence of strong prior information, one may wish to adopt a soft initial outlier threshold in the range $Q\sim 3$--$5$ (recall that $Q$ is roughly the scale at which departures from the model start looking like outliers, in units of standard deviations).
This default should provide a reasonable first balance between robustness and false-positive outlier identification.
(3) Choose the rank ($K$) using the held-out KL-divergence score of Eq.~\eqref{eq:score} and some measure of the goodness-of-fit like $\chi_\textrm{red}^2$, evaluated on a modest grid of $(Q, K)$ values (and potentially update $Q$), rather than selecting $K$ only after fitting as one might for PCA.
(4) Inspect the residuals ${\bf Y} - {\bf A} {\bf G}$ in conjunction with the validation score.
A well-described inlier population should leave residuals that are largely unstructured across objects.
In contrast, coherent residual structure shared across many objects may indicate the presence of additional shared structure not represented by the current model.
This can either be due to an outlier population with common low-rank structure (as discussed above), or because the chosen rank is not sufficient to describe all inlier structure.
This is largely up to interpretation and may be informed by pre-existing knowledge of the dataset: one must \emph{decide} whether to increase $K$ and absorb this structure into the model, or retain it as anomalous.
This choice is application-dependent and cannot be resolved by the model alone; it requires scientific judgement.

In many applications, little further tuning should be required for RHMF to provide a robust low-rank description of heteroskedastic data with missing entries, together with per-pixel and per-object outlier scores derived from a single objective function.

An installable version of the Robusta-HMF code in python is available at \url{https://github.com/TomHilder/robusta-hmf}, along with all of the code used to make the figures in this paper.

\begin{acknowledgments}  
  The authors warmly thank the anonymous referee, who provided a very timely and thorough review, which improved the clarity of the paper and helped demonstrate both the robustness and utility of the method.
  The authors thank the Stars group at MPIA, the Astronomical Data group at the Flatiron Institute, and the Inference group at Monash University, for valuable discussions.
  TH is supported by an Australian Government Research Training Program (RTP) Scholarship.
  The Flatiron Institute is a division of the Simons Foundation.
  DWH acknowledges that the land politically designated as New York City is the homeland of the Lenape (Lenapehoking), who were unjustly displaced.
\end{acknowledgments}

\software{
    \texttt{astropy} \citep{astropy2022},
    \texttt{equinox} \citep{kidger2021equinox},
    \texttt{h5py} \citep{collette2013h5py},
    \texttt{JAX} \citep{jax2018github},
    \texttt{matplotlib} \citep{hunter2007matplotlib},
    \texttt{numpy} \citep{harris2020numpy},
    \texttt{polars} \citep{polars},
    \texttt{scipy} \citep{virtanen2020scipy}
}

\appendix

\section{Properties of the objective function} \label{sec:proof}

In Eqns.~\eqref{eq:robust_opt} and \eqref{eq:t_like} we wrote the model as optimizing a Student-t likelihood.
The corresponding negative log-likelihood is
\begin{align}
    \mathcal{L}(A,G) = \frac{Q^2}{2} \sum_{ij} \log\left(1 + \frac{w_{ij}^{\rm data}\,r_{ij}^2}{Q^2}\right), \label{eq:t_loss}
\end{align}
which is identical to the loss function in Eq.~\eqref{eq:robust_opt} up to the constant $Q^2 / 2$, where $r_{ij} = y_{ij} - \sum_{k} a_{ik} g_{kj}$ is the residual and $w_{ij}^{\rm data}$ the measurement precision, exactly as in the main text.
In the remainder of this section, we prove that the alternating least squares plus weight-updating step algorithm does actually minimize this negative log-likelihood.
Throughout this appendix, $w_{ij} \in (0,1]$ denotes the robust weight, kept distinct from the input measurement precisions $w_{ij}^{\rm data}$.

For convenience, we will drop the $ij$ indices for the parts of the following that are general; $w^{\rm data}$ then denotes the (fixed, positive) measurement precision of the element under consideration.
We claim that the loss Eq.~\eqref{eq:t_loss} for any $r$ can be expressed in the following way
\begin{equation}
    \rho(r) = \min_{0<w\leq1} \left[ \tfrac{1}{2} w\, w^{\rm data} r^2 + \phi(w) \right], \label{eq:rho_loss}
\end{equation}
where
\begin{equation}
    \phi(w) = \frac{Q^2}{2}\,(w - 1 - \log w).
\end{equation}
That is, performing the minimization in Eq.~\eqref{eq:rho_loss} is identical to evaluating Eq.~\eqref{eq:t_loss}.

We will now prove the claim. First, define
\begin{align}
    J\left( w ; r \right) &= \tfrac{1}{2} w\, w^{\rm data} r^2 + \tfrac{Q^2}{2} \left( w - 1 - \log{w}\right), \label{eq:J_def}
\end{align}
where the $;$ notation just indicates that $J$ is a function of $w$ given a particular $r$.
Differentiating with respect to $w$ gives
\begin{equation}
    \frac{\partial J}{\partial w}
    = \tfrac{1}{2}w^{\rm data} r^2 + \frac{Q^2}{2}\left(1 - \frac{1}{w}\right),
\end{equation}
and again
\begin{align}
    \frac{\partial^2 J}{\partial w^2} = \frac{Q^2}{2} \frac{1}{w^2} > 0, \qquad \forall \, Q,w > 0.
\end{align}
Since the second derivative is guaranteed positive, we know that the critical point in $w$ minimises $J$.
Setting $\partial_w J=0$ to find the critical point $\hat{w}$ yields
\begin{align}
    \hat{w}(r) &= {\rm argmin}_w \, J (w ; r) \\
    &= \frac{1}{1 + w^{\rm data} r^2/Q^2}, \label{eq:w_hat}
\end{align}
and note that $\hat{w} \in (0, 1]$ because $w^{\rm data} r^2/Q^2 \geq 0$.
Now, we substitute $t = w^{\rm data} r^2/Q^2 \geq 0$, such that
\begin{equation}
    \hat{w} = \frac{1}{1+t}, \qquad w^{\rm data} r^2 = Q^2 t,
\end{equation}
and then substitute $\hat{w}$ into Eq.~\eqref{eq:J_def} to do the minimization in Eq.~\eqref{eq:rho_loss}
\begin{align}
    J (\hat{w} ; r) &= \tfrac{1}{2} \hat{w}\, w^{\rm data} r^2 + \phi(\hat{w}),
\end{align}
substituting and simplifying one piece at a time:
\begin{align}
    \Rightarrow \tfrac{1}{2} \hat{w}\, w^{\rm data} r^2 &= \frac{Q^2}{2} \frac{t}{1 + t} \\
    \Rightarrow \phi(\hat{w}) &= \frac{Q^2}{2} \left( \hat{w} - 1 - \log{\hat{w}} \right) \\
    &= \frac{Q^2}{2} \left( -\frac{t}{1 + t} + \log(1 + t) \right) \\
    \Rightarrow J (\hat{w} ; r) &= \frac{Q^2}{2} \log{\left( 1 + \frac{w^{\rm data} r^2}{Q^2} \right)}.
\end{align}
Thus
\begin{align}
    \rho(r) &= J ( \hat{w} ; r ) \\
    &= \min_{0<w\leq1} \left[ \tfrac{1}{2} w\, w^{\rm data} r^2 + \phi(w) \right],
\end{align}
as claimed.

To see how this naturally leads to the 3-step algorithm outlined in Section~\ref{sec:fitting}, we first define a new objective:
\begin{equation}
    J(A,G,W) = \frac{1}{2}\sum_{ij} \left[ w_{ij}\, w_{ij}^{\rm data}\, r_{ij}^2 + \phi(w_{ij}) \right],
\end{equation}
where $[A]_{ij} = a_{ij}$, $[G]_{ij} = g_{ij}$, $[W]_{ij} = w_{ij}$, and $r_{ij} = y_{ij} - \sum_{k} a_{ik} g_{kj}$ is the residual as in Eq.~\eqref{eq:t_loss}, and where we have reintroduced the $ij$ indices.
By construction,
\begin{equation}
    L(A,G) = \min_W J(A,G,W),
\end{equation}
and if
\begin{align}
    \hat{W} &= {\rm argmin}_W \, J(A, G, W),
\end{align}
then
\begin{align}
    \left[ \hat{W} \right]_{ij} &= \hat{w} (r_{ij}) \\
    &= \frac{1}{1 + w_{ij}^{\rm data}\, r_{ij}^2 / Q^2}, \\
    &= \frac{Q^2}{Q^2 + w_{ij}^{\rm data}\, r_{ij}^2}.
\end{align}
The optimizer of the free variable $w_{ij}$ is therefore exactly the robust weight $w_{ij}^{\rm robust}$ of Eq.~\eqref{eq:w_robust}.
This immediately yields the three-step procedure (the rotation isn't relevant for convergence)
\begin{align*}
    \text{w-step:} \quad & w_{ij} \leftarrow \hat{w}(r_{ij}), \\
    \text{a-step:} \quad & \text{solve WLS for $A$ with new weights}, \\
    \text{g-step:} \quad & \text{solve WLS for $G$ with new weights}.
\end{align*}
where the a-step optimises the quadratic
\begin{align}
    \mathcal{Q}(A \, | \, G, W) &= \frac{1}{2} \sum_{ij} w_{ij}\, w_{ij}^{\rm data}\, r_{ij}^2,
\end{align}
and the g-step optimises $\mathcal{Q}(G \, | \, A, W)$.
The least-squares solves therefore use the combined weight $w_{ij}\, w_{ij}^{\rm data} = w_{ij}^{\rm (new)}$, matching the algorithm of Section~\ref{sec:fitting}.

More explicitly, consider one outer cycle starting at $(A^{(t)}, G^{(t)})$.
Choose $W^{(t)} = \hat{w}(r(A^{(t)},G^{(t)}))$.
Then
\begin{equation}
    L(A^{(t)}, G^{(t)}) = J(A^{(t)}, G^{(t)}, W^{(t)}).
\end{equation}
With frozen $W^{(t)}$, the a- and g-steps minimize $\mathcal{Q}(\cdot \,| \,W^{(t)})$.
Since our total objective is $J = \mathcal{Q} + \sum \phi(W^{(t)})$, this implies
\begin{equation}
    J(A^{(t+1)}, G^{(t+1)}, W^{(t)}) \le J(A^{(t)}, G^{(t)}, W^{(t)}).
\end{equation}
We're guaranteed to be helped by the w-step again now, so setting
\begin{align}
    W^{(t+1)} = \hat{w} \left( r(A^{(t+1)}, G^{(t+1)}) \right),
\end{align}
and using our result from the previous section gives
\begin{equation}
    J(A^{(t+1)}, G^{(t+1)}, W^{(t+1)}) \le J(A^{(t+1)}, G^{(t+1)}, W^{(t)}).
\end{equation}
Thus chaining the inequalities and $L(A,G) = \min_W J(A, G, W)$ gives
\begin{equation}
    L(A^{(t+1)}, G^{(t+1)}) \le L(A^{(t)}, G^{(t)}).
\end{equation}
This guarantees that robust HMF with the iterative procedure converges to the Student-t maximum likelihood estimate.

\section{Toy Dataset} \label{sec:toy_model_details}

The toy dataset consisted of $N=8000$ synthetic (toy) spectra, on an $M=1200$ pixel grid, logarithmically spaced between 400 and 600 nm.
Specifically, the pixel grid was generated using the \texttt{NumPy} function \texttt{geomspace(400, 600, 1200)}.
The spectra were convolved with a Gaussian kernel to achieve a constant spectral resolution of $R = 2123$, where we actually chose the standard deviation to be $\sigma = \bar{\lambda} / 5000$, where $\bar{\lambda}$ is the mean wavelength.

The non-outlier spectra were generated from the sum of 5 basis functions (and so have a true rank of 5), according to the following
\begin{align}
    \mathrm{spectrum}_{i}(j) &= a_i + b_i z_j + c_i \left[\frac{1}{2} \left( 3 z_j^2 - 1 \right)\right] - 0.45 g_i \sum_{\mu \in \mathcal{C}} \exp{\left[ - \frac{1}{2} \left( \frac{\lambda_j - \mu}{1.4~\mathrm{nm}} \right)^2  \right]} + s_i \sin{\left( 2 \pi \left( z_j + 1 \right)  \right)},
\end{align}
where the rescaled coordinate $z_j\in[-1, 1]$
\begin{align}
    z_j &= \frac{2 \left( \log{\lambda_j} - \log{400~\mathrm{nm}} \right)}{\left( \log{600~\mathrm{nm}} - \log{400~\mathrm{nm}} \right)} - 1,
\end{align}
is such that the first three basis functions are Legendre polynomials in $\log{\lambda}$ (these functions are orthogonal in the interval $[-1, 1]$).
The set of line centres $\mathcal{C}$ was chosen as
\begin{align} 
    \mathcal{C} = \{\,
        & 410.17~\mathrm{nm}, \\
        & 434.05~\mathrm{nm}, \\
        & 486.13~\mathrm{nm}, \\
        & 516.73~\mathrm{nm}, \\
        & 517.27~\mathrm{nm}, \\ 
        & 518.36~\mathrm{nm}, \\
        & 589.00~\mathrm{nm}, \\
        & 589.60~\mathrm{nm} \,\}.
\end{align}
The $0.45$ prefactor in front of the absorption lines was set by hand to yield lines of a somewhat realistic looking depth.
The coefficients $a_i, b_i, c_i, g_i, s_i$ were randomly drawn for each spectrum $i$ from the following distributions
\begin{align}
    a_i &\sim \mathcal{N} \left( 1.0, 0.1^2 \right), \\
    b_i &\sim \mathrm{Exp} \left( 0.2 \right), \\
    c_i &\sim \mathcal{N} \left( 0.0, 0.1^2 \right), \\
    g_i &\sim \log{\mathcal{N}} \left( 0.0, 0.1^2 \right), \\
    s_i &\sim \mathrm{N} \left( 0.1, 0.1 \right),
\end{align}
where $\mathcal{N}(\mu, \sigma^2)$ denotes the normal distribution with mean $\mu$ and variance $\sigma^2$, $\mathrm{Exp}(s)$ denotes the exponential distribution with scale $s$, and $\log{\mathcal{N}}(\mu, \sigma^2)$ denotes a log normal distribution where the mean and standard deviation are that of the underlying normal.

Noise draws $\epsilon_{ij}$ were added to the spectra, generated as
\begin{align}
    \epsilon_{ij} \sim \mathcal{N} \left( 0, \sigma_{ij}^2\right),
\end{align}
The \emph{known}, per-pixel, Gaussian uncertainties $\sigma_{ij}$ were chosen as
\begin{align}
    \sigma_{ij} = 0.2 \, \alpha_i \, \beta_j \, \gamma_{ij},
\end{align}
with random per-spectrum multiplier
\begin{align}
    \alpha_i^* &\sim \log{\mathcal{N}} \left( 1.0, 2^2 \right), \\
    \alpha_i &= \alpha_i^* / \bar{\alpha_i^*},
\end{align}
wavelength-dependent per-pixel multiplier
\begin{align}
    \beta_j &= \exp{\left[ - \frac{1}{2} \left( \frac{\lambda_j - \bar{\lambda}}{100~\mathrm{nm}} \right)^2  \right]},
\end{align}
and random per-spectrum-per-pixel jitter multiplier
\begin{align}
    \gamma_{ij}^* &\sim \log{\mathcal{N}} \left( 0, 0.06^2 \right), \\
    \gamma_{ij} &= \gamma_{ij}^* / \bar{\gamma_{ij}^*},
\end{align}
where in all cases the hand-tuned numbers above were set to give a sensible range by eye, and the bar notation $\bar{x}$ denotes the mean.

A randomly selected 40 of the 8000 spectra were replaced with the following outlier model
\begin{align}
    \mathrm{outlier}_{i}(j) = 1 + A_i \sin{\left( 2\pi f_i \left[\frac{j}{M-1}\right] \right)},
\end{align}
which notably has constant frequency $f_i$ in the pixel index $j$, not in the wavelength.
The amplitudes and frequencies were randomly drawn as
\begin{align}
    A_i &\sim \mathcal{U} \left( 0.1, 0.3 \right), \\
    f_i &\sim \mathcal{U} \left( 10, 30 \right),
\end{align}
where $\mathcal{U} \left( l, r \right)$ denotes a bounded uniform distribution over the interval $[l,\,r]$.
These outliers share effectively no common low-rank structure amongst themselves nor with the spectrum model.

Outlier ``columns'' were injected such that for 3 randomly selected pixel indices $j$, 30\% of all the spectra (selected randomly) had a factor $\delta_{ij} \sim \pm\,\mathcal{U} \left( 0.25, 0.45 \right)$ added in pixel $ij$, where the sign was also random.
Individual outlier pixels were also added in 0.4\% of all pixels, for a total of 38,400.
The pixel and spectrum indices were both drawn randomly, according to $\delta_{ij} \sim \pm\,\mathcal{U} \left( 0.25, 0.75 \right)$, where again the sign was random.
Both column and pixel outliers were injected across all spectra, and so could affect both the normal and outlier spectra.

10 normal (non-outlier) spectra also had 3 extra absorption lines added.
Note that these spectra could also contain column and pixel outliers, and we did not prevent any of them from being coincident in location.
Each of these were centred on a randomly (and uniformly) selected pixel, with a width (standard deviation) of 2 pixels, and a depth drawn as $d \sim \mathcal{U} \left( 0.3, 0.6 \right)$.
Each of the 10 spectra had \emph{different} sets of lines added, in that the random sampling was done for each of the 30 lines.

Continuous missing data segments were injected in 50\% of all spectra, selected randomly.
The length of the segment was drawn uniformly between 50 and 200.
The locations were also chosen randomly, except that overlap with pre-existing pixel outliers and the extra absorption lines was avoided.
They were however, allowed to overlap with the column outliers, and to be placed in outlier spectra.
Missing values were set to \texttt{NaN}, and the corresponding data weights were set to zero.

\begin{figure}
    \centering
    \includegraphics[width=\textwidth]{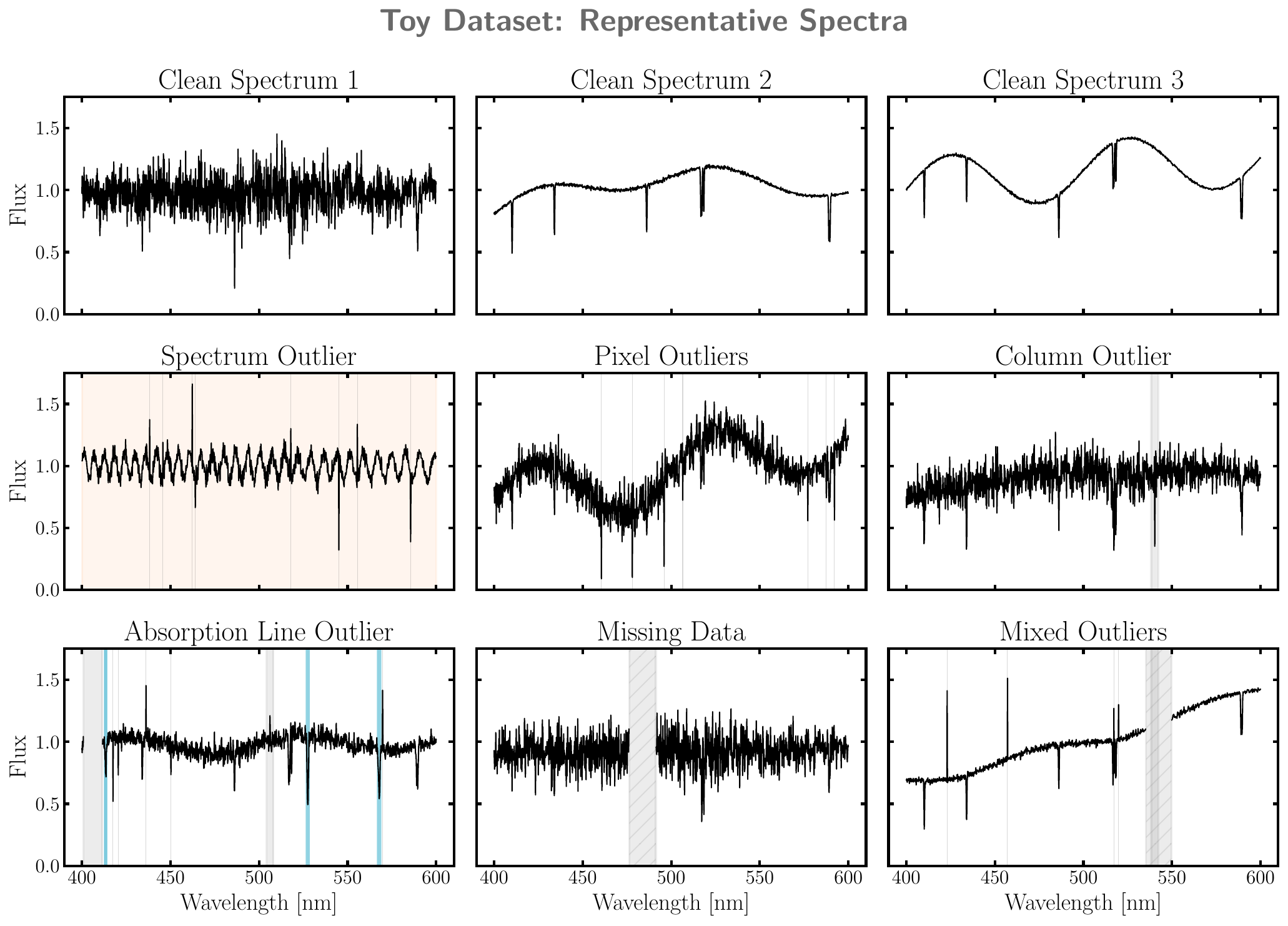}
    \caption{Representative spectra from the toy dataset illustrating dataset diversity. Top row: normal spectra (with only Gaussian noise), showing the range of underlying spectral shapes. Middle row: individual outlier types—spectrum-level (high-frequency sinusoid), pixel-level (scattered spikes), and column-level (same wavelength affected consistently). Bottom row: absorption-line outliers with extra spectral features, missing data segments (visible as gaps in the spectrum), and mixed outlier types. Outlier regions are highlighted with low-alpha colored shading; solid colored ticks mark individual anomalies.}
    \label{fig:toy_dataset_diversity}
\end{figure}

\begin{figure}
    \centering
    \includegraphics[width=\textwidth]{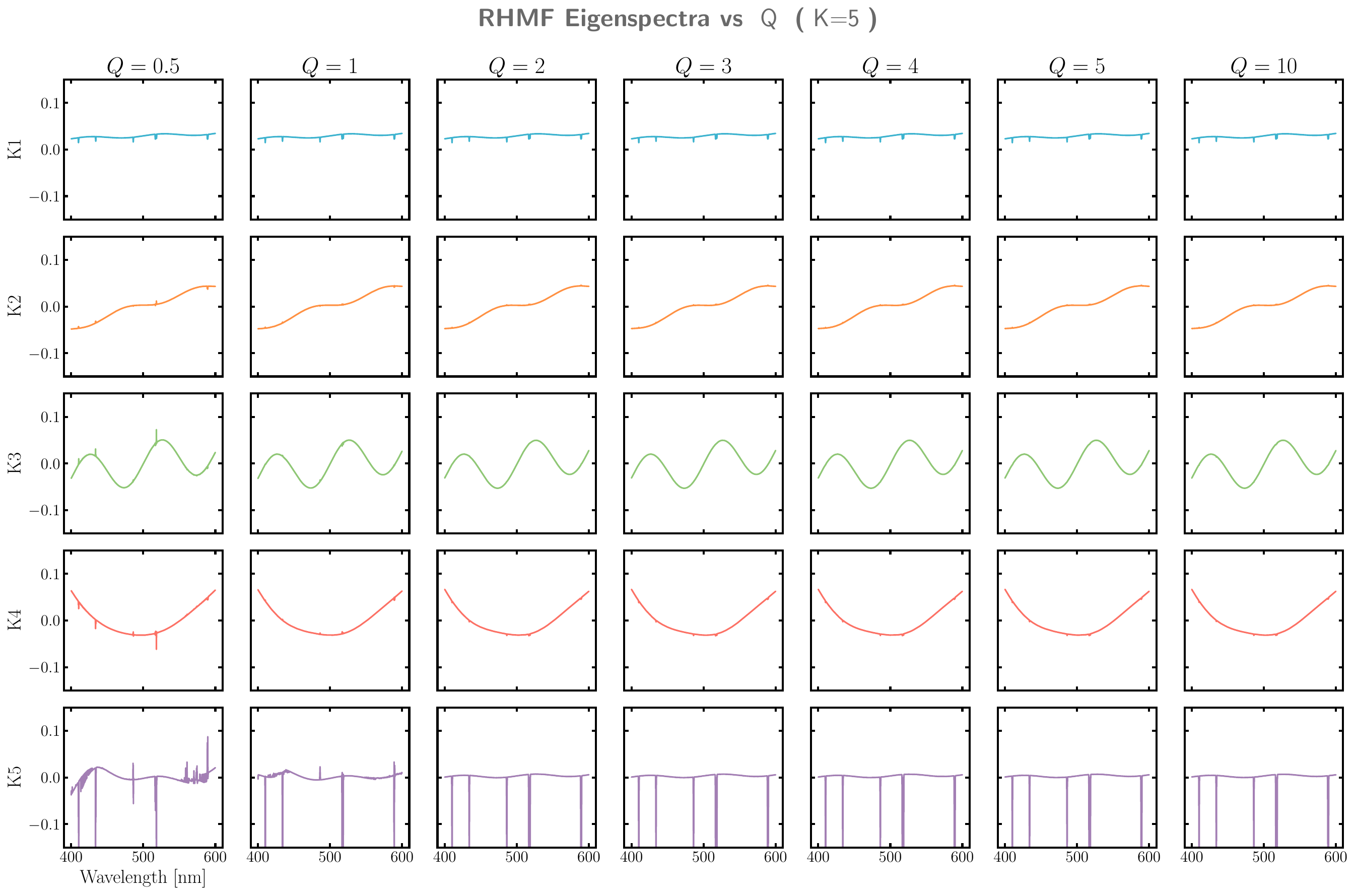}
    \caption{RHMF eigenspectra as a function of the robust scale $Q$ at the true rank $K=5$, in the canonical frame of Figure~\ref{fig:eigenspectra}. The recovered basis is essentially invariant for $Q \geq 2$; at $Q \lesssim 1$ the aggressive downweighting begins to erode the weakest component (K5, the absorption-line template).}
    \label{fig:eigenspectra_vs_q}
\end{figure}

\begin{figure}
    \centering
    \includegraphics[width=0.9\textwidth]{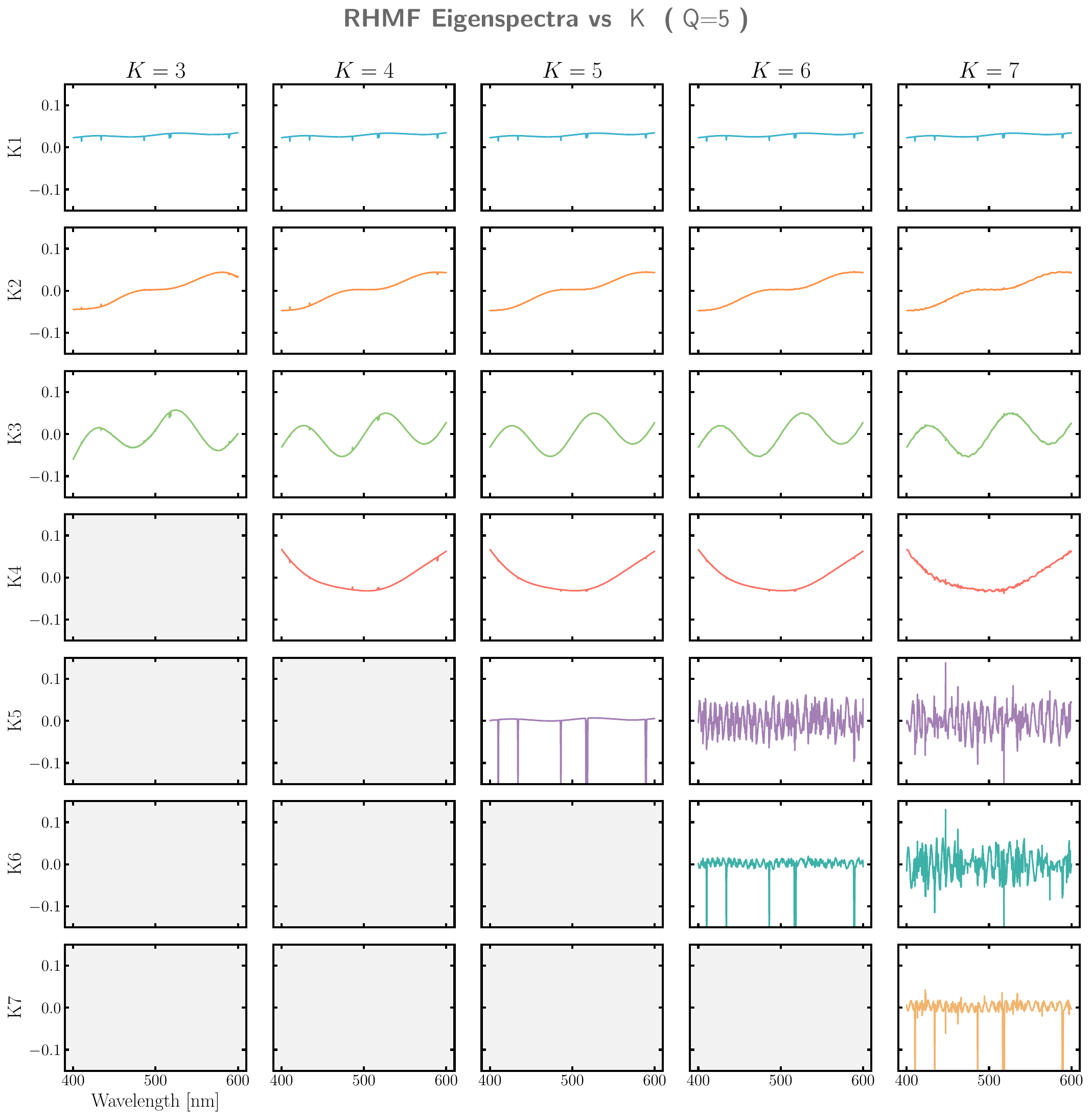}
    \caption{RHMF eigenspectra as a function of the rank $K$ at the CV-optimal robust scale $Q=5$. Greyed panels indicate components that do not exist for that model. Components enter in order of their data-power contribution as $K$ increases; for $K$ above the true rank the leading components are unchanged and the extra components absorb noise.}
    \label{fig:eigenspectra_vs_k}
\end{figure}

Figures~\ref{fig:eigenspectra_vs_q} and \ref{fig:eigenspectra_vs_k} show how the recovered RHMF eigenspectra vary across the $(K, Q)$ hyperparameter grid of Section~\ref{sec:toy}, with every model expressed in the canonical frame of Figure~\ref{fig:eigenspectra}. The basis is essentially unchanged from $Q=2$ to $Q=10$ at the true rank; only for very aggressive downweighting ($Q \lesssim 1$) does the weakest component (the absorption-line template) degrade, as genuine line pixels begin to be treated as outliers. Varying $K$ at fixed $Q=5$, components enter in order of their data-power contribution: $K=3$ recovers the three strongest components, $K=4$ adds the quadratic continuum mode, and $K=5$ adds the line template. For over-parameterized models ($K = 6, 7$) the leading components are unchanged and the extra components absorb noise, with the line template mixing among the near-degenerate trailing components. This illustrates, at the level of the basis vectors themselves, the hyperparameter insensitivity discussed in Section~\ref{sec:toy}: results are stable provided $K$ is at least the true rank and $Q$ is not extreme.

\section{Gallery of outlier classes in the full Gaia RVS sample} \label{sec:tax_gallery}

\begin{figure}
    \centering
    \includegraphics[width=0.9\textwidth]{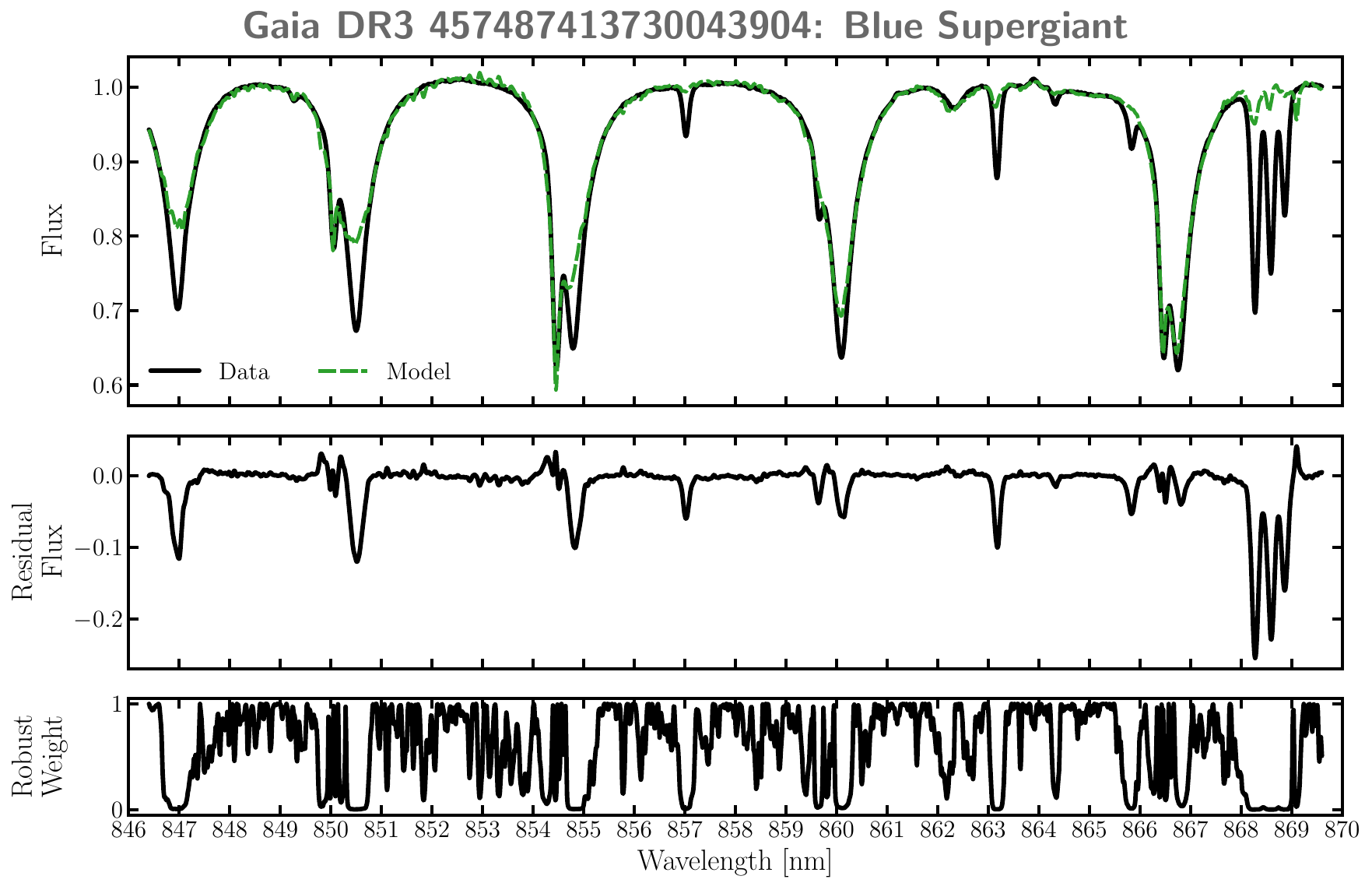}
    \caption{The most anomalous star identified by our model in Gaia/RVS spectra (Gaia DR3 457487413730043904), a known blue supergiant. Many similar examples of blue supergiants are identified by their low median spectrum weights.}
    \label{fig:tax_bsg}
\end{figure}

\begin{figure}
    \centering
    \includegraphics[width=0.9\textwidth]{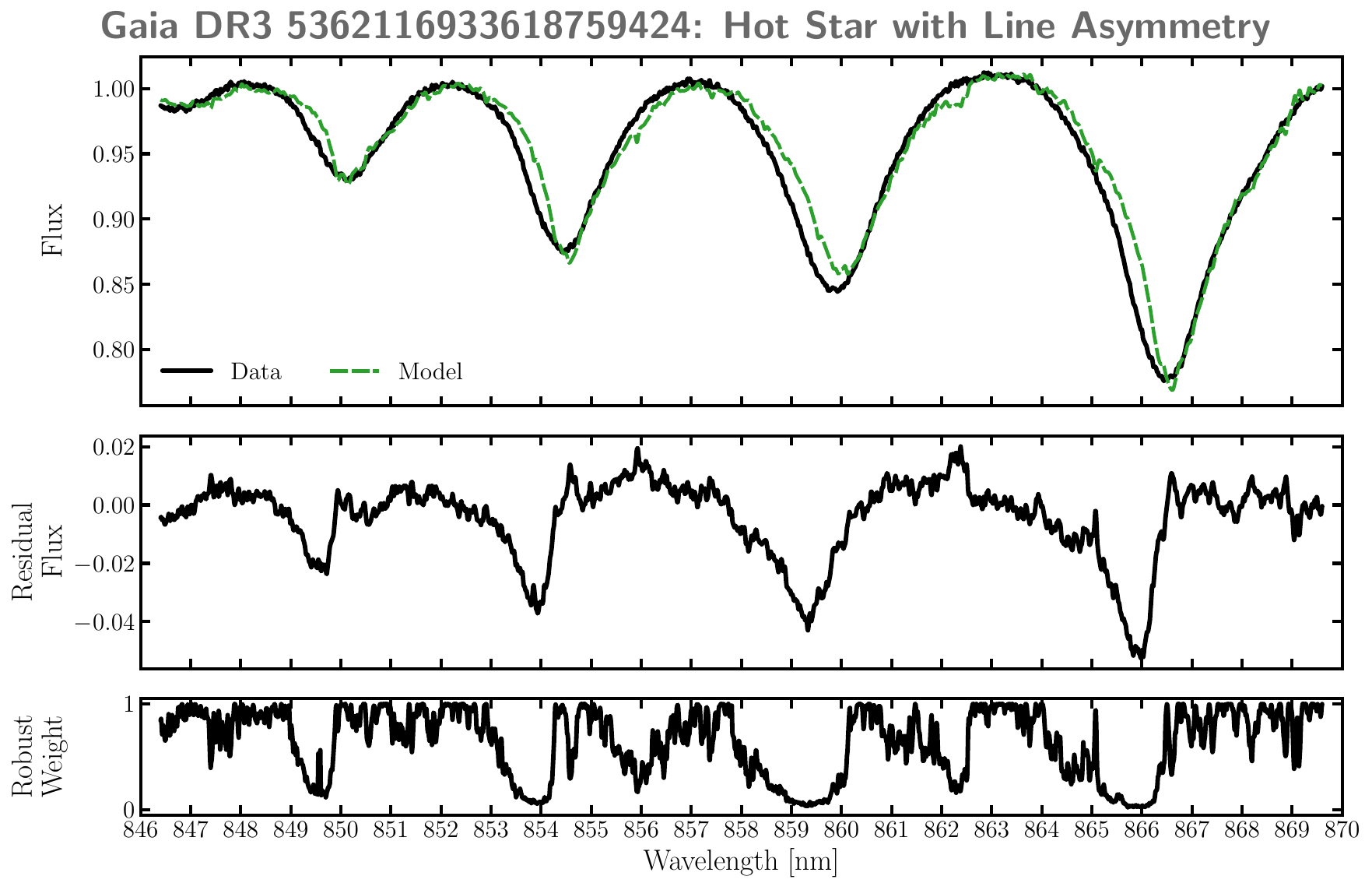}
    \caption{A hot star with asymmetric line profiles (Gaia DR3 5362116933618759424). The data show a coherent blueward shift relative to the model at the Paschen lines.}
    \label{fig:tax_hot_asymmetric}
\end{figure}

\begin{figure}
    \centering
    \includegraphics[width=0.9\textwidth]{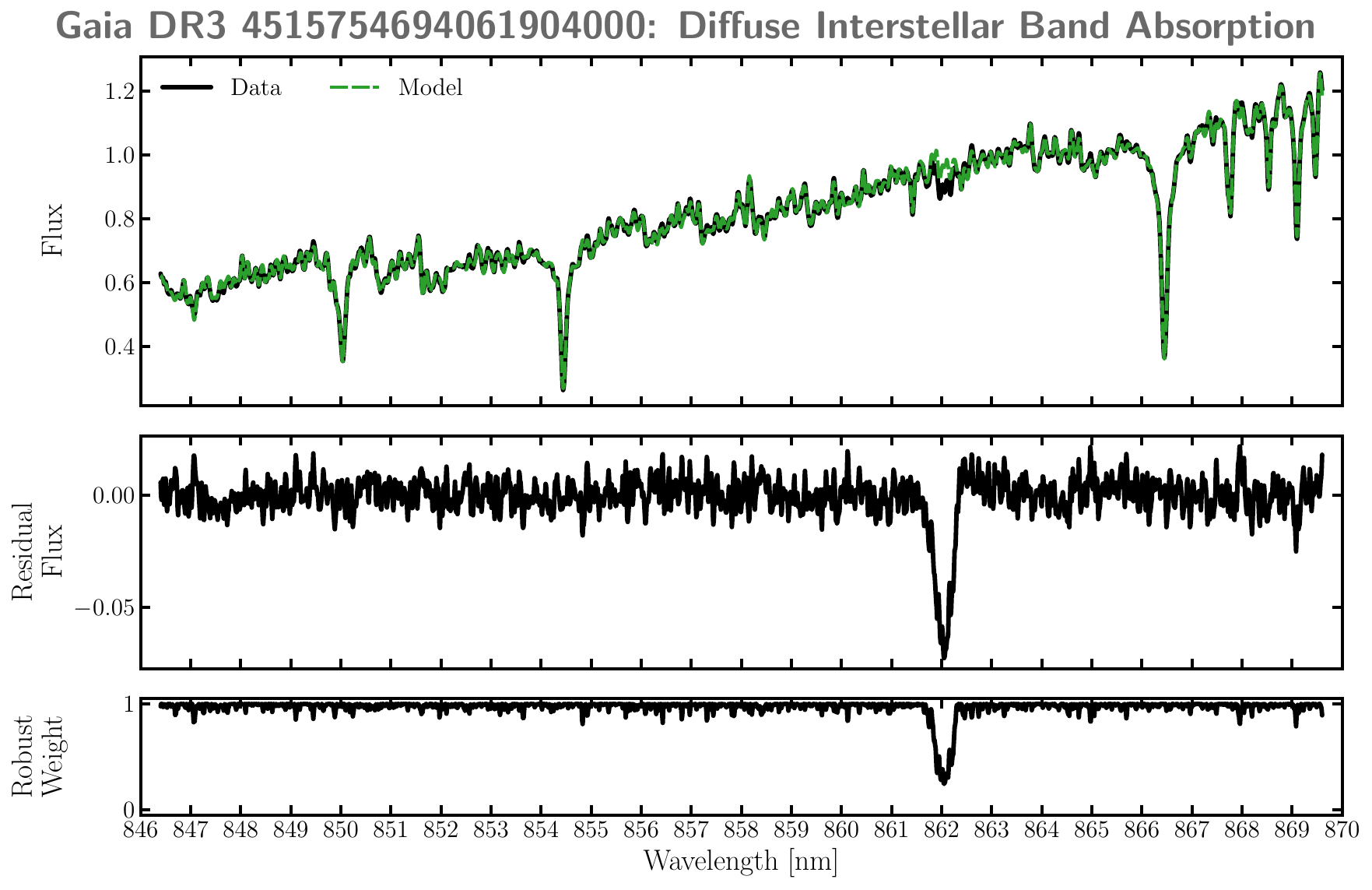}
    \caption{A star with strong diffuse interstellar band absorption (Gaia DR3 4515754694061904000). The residual is dominated by the 862~nm diffuse interstellar band, which is cleanly downweighted while the rest of the spectrum is well fit by the model.}
    \label{fig:tax_dib}
\end{figure}

\begin{figure}
    \centering
    \includegraphics[width=0.9\textwidth]{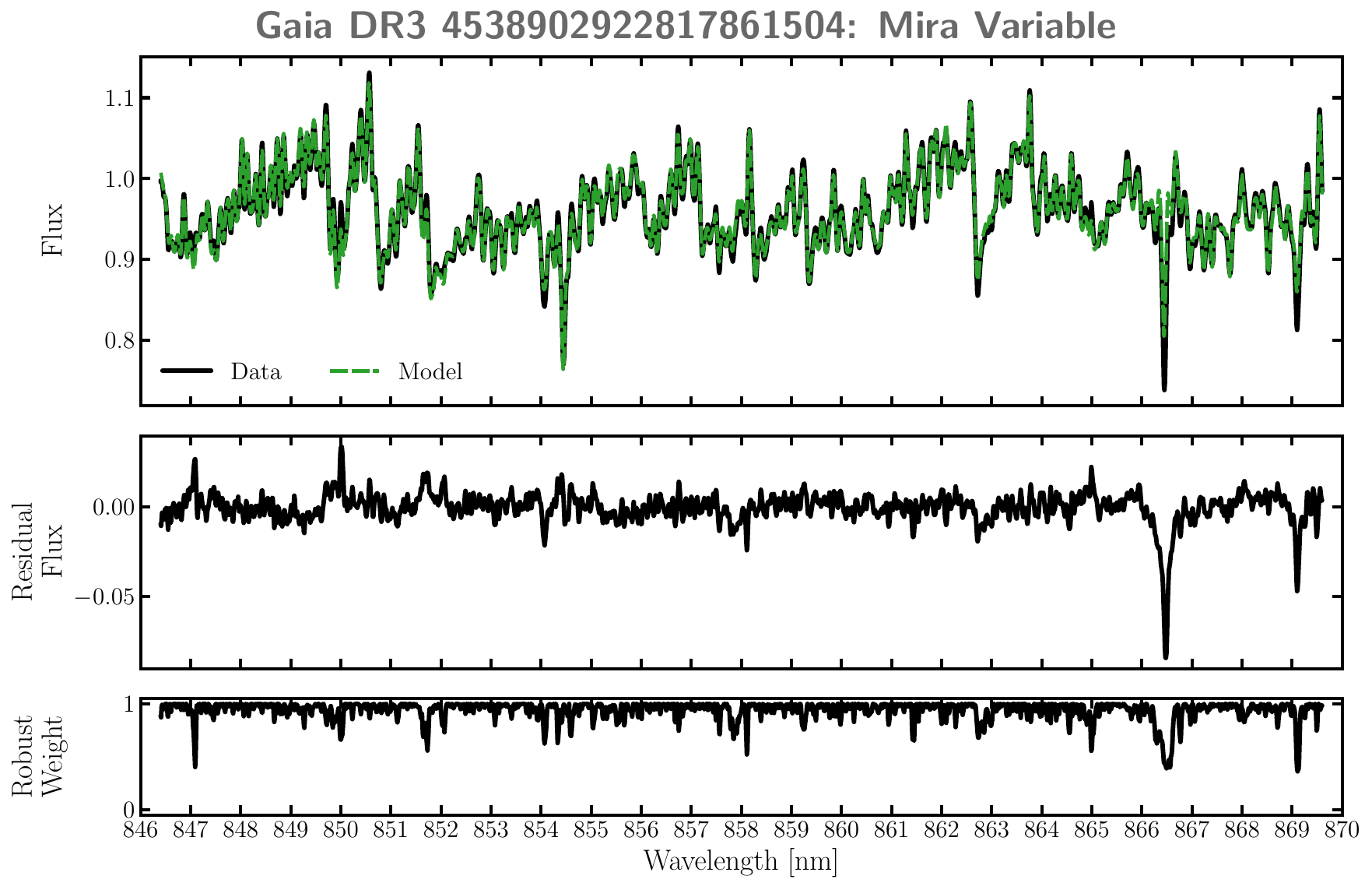}
    \caption{A known Mira variable (Gaia DR3 4538902922817861504) that is identified as an outlier by its low median spectrum weights. The molecular-band-dominated spectrum is broadly reproduced, with localised residuals near the strongest features.}
    \label{fig:tax_mira}
\end{figure}

\begin{figure}
    \centering
    \includegraphics[width=0.9\textwidth]{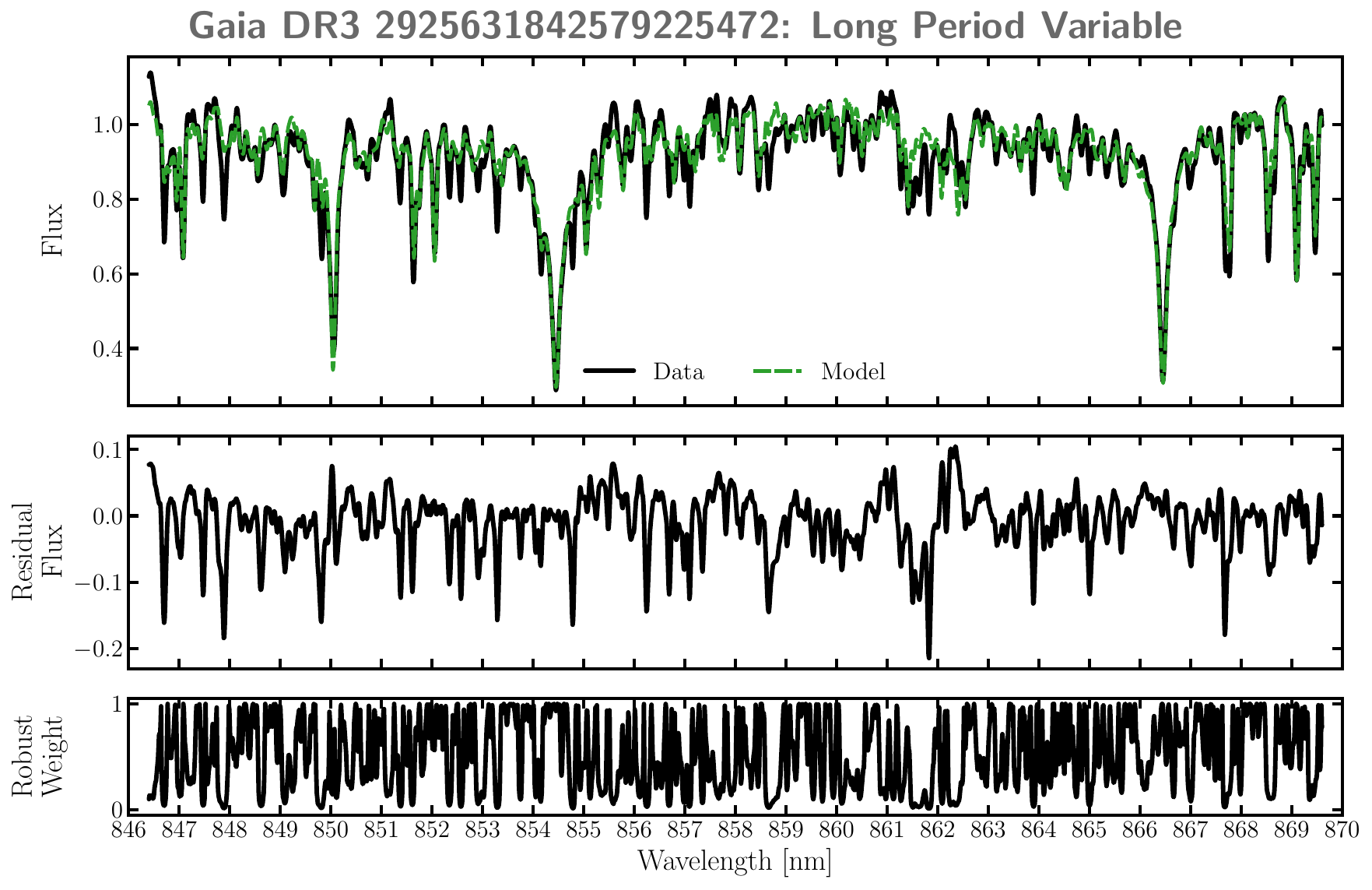}
    \caption{A known long period variable (Gaia DR3 2925631842579225472). The dense forest of molecular features is poorly captured by the basis, leading to widespread downweighting.}
    \label{fig:tax_lpv}
\end{figure}

\begin{figure}
    \centering
    \includegraphics[width=0.9\textwidth]{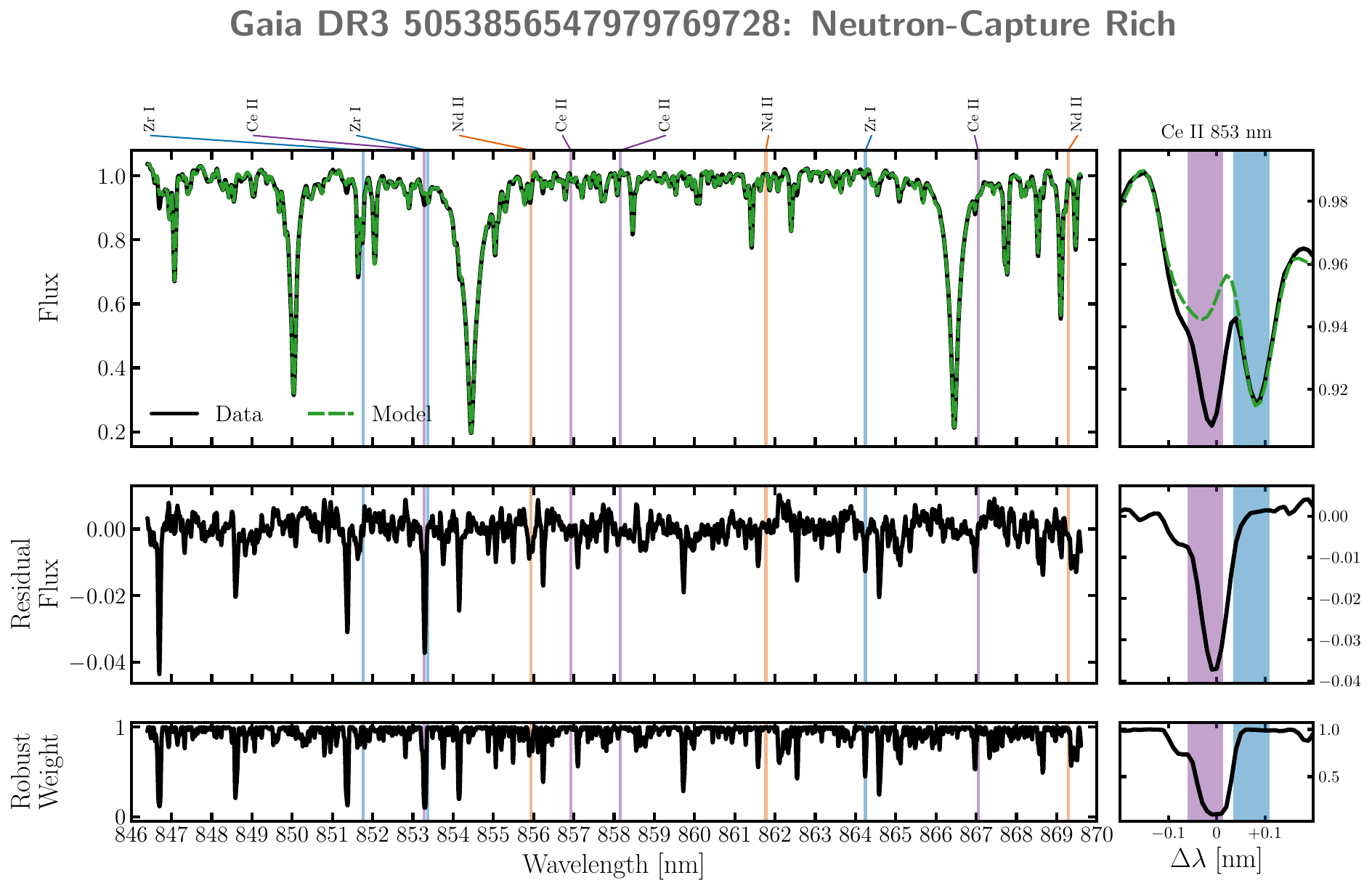}
    \caption{A star seemingly abundant in the neutron-capture element cerium (Gaia DR3 5053856547979769728). Other neutron-capture absorption lines are highlighted, although most are very weak. Inset panels on the right highlight the region around the Ce II line at 853 nm.}
    \label{fig:tax_ncap}
\end{figure}

Figures~\ref{fig:tax_bsg} to \ref{fig:tax_ncap} show representative examples of some of the  frequent outlier classes identified by visual inspection of the $1{,}559$ spectra flagged in Section~\ref{sec:gaia_full}. Each figure shows the observed spectrum and the RHMF reconstruction (top), the residual flux (middle), and the per-pixel robust weights (bottom).

\raggedright\footnotesize
\bibliographystyle{aasjournal}
\bibliography{rhmf}

\end{document}